\documentclass[aps,prb,10pt,twocolumn,longbibliography]{revtex4-1}
\usepackage{amsmath}
\usepackage{amsfonts}
\usepackage{amssymb}
\usepackage{graphicx}
\usepackage{bm}
\usepackage[colorlinks=true,linktocpage=true,breaklinks=true]{hyperref}

\begin{document}

\title{Spin-orbit torque in 3D topological insulator-ferromagnet heterostructure:\\ crossover between bulk and surface transport}
\author{S. Ghosh}
\email{sumit.ghosh@kaust.edu.sa}
\author{A. Manchon}
\email{aurelien.manchon@kaust.edu.sa}
\affiliation{King Abdullah University of Science and Technology (KAUST), Physical Science and Engineering Division (PSE), Thuwal 23955, Saudi Arabia}

\begin{abstract}
Current-driven spin-orbit torques are investigated in a heterostructure composed of a ferromagnet deposited on top of a three dimensional topological insulator using the linear response formalism. We develop a tight-binding model of the heterostructure adopting a minimal interfacial hybridization scheme that promotes induced magnetic exchange on the topological surface states, as well as induced Rashba-like spin-orbit coupling in the ferromagnet. Therefore, our model accounts for spin Hall effect from bulk states together with inverse spin galvanic and magnetoelectric effects at the interface {\em on equal footing}. By varying the transport energy across the band structure, we uncover a crossover from surface-dominated to bulk-dominated transport regimes. We show that the spin density profile and the nature of the spin-orbit torques differ substantially in both regimes. Our results, which compare favorably with experimental observations, demonstrate that the large damping-like torque reported recently is more likely attributed to the Berry curvature of interfacial states, while spin Hall torque remains small even in the bulk-dominated regime.\end{abstract}

\maketitle

%%%%%%%%%%%%%%%%%%%%%%%%%%%%%%%%%%%%%%%%%%%
\section{Introduction}

The conventional way to record information on magnetic elements exploits spin transfer torque (STT), a mechanism that transfers spin angular momentum from a $reference$ to a $free$ magnetic layer via a spin polarized current \cite{Brataas2012}. This effect has been used successfully in the context of magnetic random access memories \cite{Kent2015}, offering both scalability and fast switching rate. Yet, its efficiency remains limited by the polarization of the reference layer and reading/writing operations must be performed through the same channel. These limitations are overcome in a spin-orbit torque (SOT) device, where the role of the polarizer is replaced by the spin-orbit coupling (SOC) of the material \cite{Bernevig2005, Manchon2008, Manchon2009, Garate2009, Brataas2014}. Since its first observation in (Ga,Mn)As \cite{Chernyshov2008}, SOT has drawn significant attention for its ability to promote fast switching \cite{MihaiMiron2010,Liu2012,Garello2014}, very high domain wall motion \cite{Miron2011b,Emori2013,Yang2015} and GHz excitations \cite{Liu2012a, Demidov2012}. Several setups have been proposed to improve the device efficiency \cite{ Fukami2016, Fukami2016a, Lau2016, VandenBrink2016, Qiu2016}. Most SOT devices are made of a magnetic layer (FM) deposited on a substrate with strong SOC. The SOC in the substrate generates a non-equilibrium spin density that exerts a torque on the adjacent FM layer and thus can manipulate its order parameter. Two types of torques are usually observed in such systems \cite{Garello2013, Qiu2015b}: (i) a field-like torque, ${\bf T}_F\sim {\bf m}\times({\bf z}\times{\bf j}_e)$, and (ii) an (anti)damping-like torque, ${\bf T}_D\sim {\bf m}\times[({\bf z}\times{\bf j}_e)\times{\bf m}]$. Here, ${\bf m}$ is the magnetization direction of the magnet, ${\bf z}$ is the normal to the interface and ${\bf j}_e$ is the injected current. Two main effects are usually considered to be at the origin of the SOT: the spin Hall effect (SHE) taking place in the bulk of the substrate induces (mostly) a damping-like torque \cite{Ando2008,Haney2013,Liu2011c} while the inverse spin galvanic effect (ISGE - also called Rashba-Edelstein effect \cite{Ivchenko1978, Edelstein1990}) present at the interface produces (mostly) a field-like torque \cite{Manchon2008,Manchon2009, Bernevig2005,Garate2009,Haney2013a}. However, several experimental \cite{Fan2013, Pai2014, Kim2013a, Kim2014a, Nguyen2016} and theoretical \cite{Freimuth2014} clues indicate that the debate is not settled yet and additional mechanisms, such as intrinsic magnetoelectric effect \cite{Kurebayashi2014, Manchon2014, Li2015} and spin swapping \cite{Lifshits2009,Saidaoui2016}, have been suggested to take place in ultrathin magnetic multilayers.
 
Due to their inherent strong SOC, topological insulators display large spin-charge conversion efficiency \cite{Shiomi2014, Jamali2015, Rojas-Sanchez2016} and are now viewed as suitable candidates for spintronics applications \cite{Fan2016}. A three-dimensional time-reversal symmetric topological insulator (TI) is characterized by an insulating bulk and surface spin-momentum locked single Dirac cones with well-defined chirality \cite{Qi2011}. Recently TIs have been found to be a powerful source of SOT \cite{Mellnik2014, Wang2015, Fan2014, Fan2015,Yasuda2017} that can be further controlled by gate fields\cite{Fan2015}. Most importantly for memory and logic applications, switching current as low as $\sim 10^5~ \rm A/cm^{2}$ at room temperature \cite{Han2017,Wang2017d,Mahendra2017} has been reported in Bi$_2$Se$_3$-based bilayers, two to three orders of magnitude smaller than their heavy metal counterpart \cite{Miron2011,Liu2012}. The parameter conventionally used to evaluate the SOT efficiency in experiments is the dimensionless {\it ``effective spin Hall angle"}. This angle (expressed in percent) quantifies the overall efficiency of the spin-charge conversion processes taking place in the heterostructure. In FM-TI heterostructures a gigantic effective spin Hall angle ranging from 160\% \cite{Han2017} to 42,500\% \cite{Fan2014} has been reported.\par

Despite these experimental breakthroughs, the physical origin of such enormous efficiencies is still under debate. While it is widely accepted that the spin-momentum locking of the surface states plays an important role, the contribution of bulk states through SHE has not been unequivocally ruled out. Till now, theories have either considered the simplistic two-dimensional Dirac Hamiltonian \cite{Garate2010, Sakai2014,Ho2016,Ndiaye2017,Fischer2016} that neglects the contributions of bulk states, or a quasi-three dimensional heterostructure that disregards intrinsic Berry-curvature induced properties \cite{Mahfouzi2014, Mahfouzi2016}. In spite of these substantial efforts, the actual physical mechanism leading to the huge SOT efficiencies reported experimentally remains to be identified unambiguously. \par

What makes FM-TI heterostructures unique and subtle is the major role played by the topological characteristics of the bulk states' wave functions \cite{Fu2007,Roy2009a, Peng2016}. Although transport through these bulk states has been neglected in the above theories, experimental evidence shows that realistic TIs are conductive and suggests that such states could substantially contribute to SOT. Another subtlety comes from the very nature of the orbital hybridization between FM and TI layers. Density functional theory reveals that not only the spin texture gets modified in presence of hybridization with a magnetic overlayer \cite{Tejada2017}, but the TI bands are also pushed below Fermi level \cite{Zhang2016,Hsu2017}, which can significantly modify the features of SOT. Till now, there is no clear explanation of how these modifications of band structure and spin texture affect the spin density and SOT. A proper description of the SOT efficiency in surface-dominated and bulk-dominated transport regimes is therefore highly solicited.\par

In the present work, we build a tight-binding model for the FM-TI heterostructure that describes bulk and surface spin transport on equal footing. Hence, interface-driven ISGE and bulk-driven SHE are computed simultaneously. By investigating the layer-resolved conductivity, non-equilibrium spin density and SOT over a wide range of transport energies, we uncover a crossover between interface-dominated and bulk-dominated regimes, associated with a substantial variation of the field-like and damping-like torque components. We show that the SOT is maximal when surface transport dominates, while the SHE arising from the bulk of the TI has a very small contribution to SOT, even in the bulk-dominated regime. We demonstrate that the damping-like torque arises from the Berry curvature of the interfacial states\cite{Garate2010,Kurebayashi2014} (also called magnetoelectric effect), and not from the bulk SHE.

%%%%%%%%%%%%%%%%%%%%%%%%%%%%%%%%%%%%%%%%%%%
\section{Model and Method}

\subsection{Tight Binding Model}
A TI surface state is characterized by a Dirac cone with strong spin-momentum locking and a well-defined chirality. When a FM material is brought to the vicinity of a TI surface, three main effects take place \cite{Zhang2016}: (i) The TI acquires an induced magnetization by proximity effect which opens a gap, (ii) the FM receives an induced (Rashba-like) SOC resulting in a modification of the spin texture in both layers, and (iii) the TI Dirac cone is pushed down in energy due to the increased carrier density at the surface. The tight-binding model described in the present section aims at accounting for these three effects through a simple hybridization scheme. \par

\begin{figure}[h]
\centering
\includegraphics[width=0.35\textwidth]{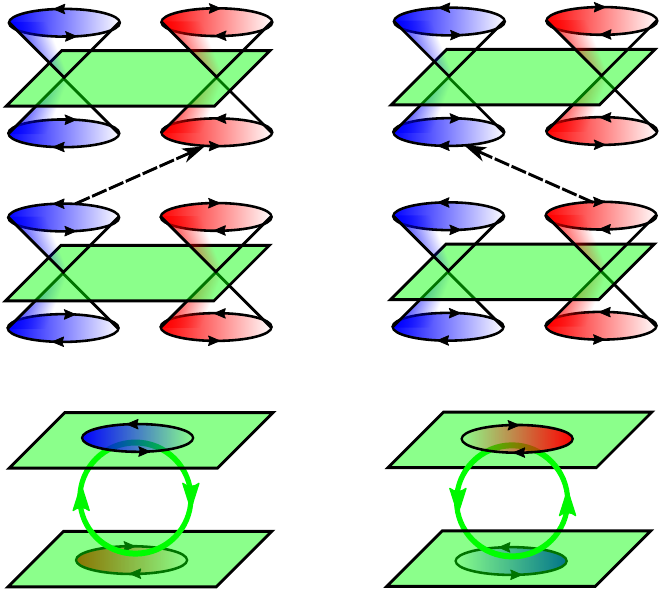}
\caption{(Color online) Schematic of the interlayer coupling procedure and origin of the bulk spin Hall current and surface texture.}
\label{coupling}
\end{figure}

We define our TI motif with a $4 \times 4$ effective Hamiltonian regularized on a cubic lattice \cite{Marchand2012}. The onsite Hamiltonian ($\hat H_0$) for each layer reads 
\begin{eqnarray}
\hat H_0 &=& \begin{pmatrix}\hat h_k + \hat u_k & \hat m_k \\ \hat m_k & -\hat h_k + \hat u_k \end{pmatrix}, \label{h0}\\
\hat h_k &=& A [\hat \sigma_y\sin (k_x a_0) - \hat{\sigma_x}\sin (k_y a_0)], \\
\hat u_k &=& \left[c- d(\cos (k_x a_0) +\cos (k_y a_0)) \right]\hat{\mathbb{I}}_2, \label{ah}\\
\hat m_k &=& \left[M- B(\cos (k_x a_0) + \cos (k_y a_0))\right] \hat{\mathbb{I}}_2,
\end{eqnarray}
where $\hat{}$ denotes an operator, $\hat{\sigma}_i$ is the $i$-th Pauli spin matrix and $\hat{\mathbb{I}}_2$ is the 2$\times$2 identity matrix. The individual layers are connected by the matrix $\hat H_T$
\begin{eqnarray}
\hat H_T&=& \begin{pmatrix}0 & \hat{\mathbb{I}}_2 (A_1-B_1)/2 \\ \hat{\mathbb{I}}_2 (-A_1-B_1)/2 & 0 \end{pmatrix}.
\end{eqnarray}
Here $A$, $B$, $M$, $c$, $d$, $A_1$ and $B_1$ are the tight binding parameters of the model and $a_0$ is the lattice constant. 
%In addition to the effective Hamiltonian defined in \cite{Marchand2012} we also introduce an additional asymmetry term $u_k$ (\ref{ah}) to make our band structure more realistic. 
The basic idea behind this model is to create a pair of chiral states at each layer and then connect each Dirac node to a state with opposite chirality in the adjacent layer in such a way that each surface only has a single Dirac node (see Fig.~\ref{coupling}). Such a scheme can be adopted to model chiral TIs and superconductors \cite{Hosur2010}. Depending on the mode of connection one can create a positive or negative spin Hall current in the bulk, which in turns determines the surface spin texture and spin density \cite{Peng2016}. %In our model this can be controlled by choosing $A_1=\pm B_1$.

The real hybridization scheme between the FM ($d$ and $p$) orbitals and the TI ($p$) orbitals is quite complex \cite{Zhang2016,Hsu2017}. In the present work, we limit ourselves to a minimal, spin-independent coupling that is sufficient to promote the effects we are interested in, i.e., induced Rashba spin splitting in FM and induced magnetization in TI. The magnetic layers are defined with a $2 \times 2$ Hamiltonian $\hat H_M$ and are coupled together by the $2 \times 2$ Hamiltonian $\hat T_M$. The coupling between the topmost TI and the bottommost FM layers is governed by a $2 \times 4$ connection matrix $\hat T_{TM}$. 
 
\begin{eqnarray}
\hat H_M &=& \left(\epsilon_0 - t_0(\cos k_x a_0+\cos k_y a_0)\right)\hat{\mathbb{I}}_2 + \Delta \hat{\sigma}_z, \label{hm}\\
\hat T_M &=& t_0 \hat{\mathbb{I}}_2, \; \hat T_{TM}= t_{\rm TM}\begin{pmatrix}1 & 0 & 1 & 0 \\ 0 & 1 & 0 & 1 \end{pmatrix}, \label{hmt}
\end{eqnarray}
where $\epsilon_0$, $t_0$ and $\Delta$ are the onsite, hopping and Zeeman energy of the FM and $t_{\rm TM}$ is the coupling between TI and FM layers. For simplicity we define the FM layer with a cubic lattice with lattice parameter $a_0$. Note that for magnetic layers we have chosen the hopping $t_0$ along the vertical direction twice stronger compared to the in-plane hopping $t_0/2$ to make sure the magnetic bands are well separated and their slope is smaller than that of the surface bands of TI. This does not change the qualitative behavior of the physical observables like spin density and makes is easier to identify the contributions of the magnetic bands when scanning through the transport energy. The complete Hamiltonian for the FM-TI heterostructure thus looks like,

\begin{eqnarray}
\hat H(k_x,k_y) &=& \begin{pmatrix}
\ddots   & \hat T_M      & 0       & 0      & 0 \\
\hat T_M^\dagger & \hat H_M      & \hat T_{TM}      & 0      & 0 \\
0      & \hat T_{TM}^\dagger & \hat H_0      & \hat H_{T}    & 0 \\
0      & 0       & \hat H_{T}^\dagger & \hat H_0     & \hat H_T\\
0      & 0       & 0       & \hat H_T^\dagger & \ddots
\end{pmatrix}.
\label{fullH}
\end{eqnarray}
%\begin{figure}[h]
%\centering
%\includegraphics[width=0.3\textwidth]{MTI}
%\caption{Schematic of MTI structure.}
%\label{blocks}
%\end{figure}

A system with $n_1$ FM layers and $n_2$ TI layers is therefore defined by a $(2n_1 + 4n_2)\times (2n_1 + 4n_2)$ matrix. For our study we choose 20 layers of TI and 5 layers of FM. We choose the parameter $A$ as our unit of energy and set the other parameters for the TI with respect to $A$ as displayed in Table~\ref{table_tb}. We keep the Zeeman splitting ($\Delta$) and coupling strength ($t_{\rm TM}$) as free parameters for now. Unless mentioned otherwise we choose $\epsilon_0$, the onsite energy of the magnetic layers, such a way that the decoupled magnetic bands starts from energy 0.3A so that we can identify the contributions coming from the FM layers and FM-TI coupling easily. The actual values of these parameters can be determined from a DFT calculation for the bulk material. For example in case of Bi$_2$Se$_3$ the bulk band gap is $\sim 0.4$ eV which can be obtained with $A$=0.27 eV. 
\begin{table}[h]
\begin{tabular}{|c|c|c|c|c|c|c|c|c|}
%\begin{tabular}{ccccccccc}
\hline 
A & B  & M & c  & d  & $A_1$ & $B_1$ & $t_0$ & $t_M$ \\
\hline
1 & 1.5 & 3.5 & 1.5 & 0.75 & 1.5  & 1.5  & 0.5  & 0.5  \\
\hline
\end{tabular}
\caption{Tight-binding parameters used in the main text.}
\label{table_tb}
\end{table}

\begin{figure}[h]
\centering
\includegraphics[width=0.45\textwidth]{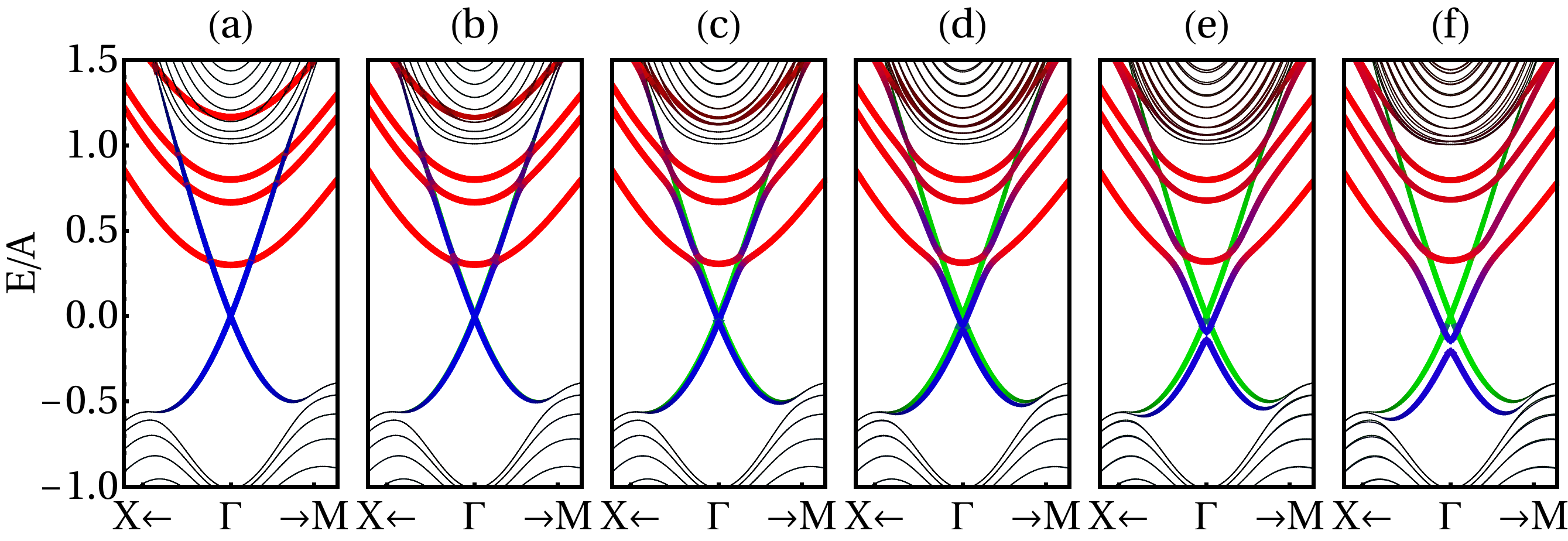}
\caption{(Color online) Band structure for 5 FM + 20 TI with $\Delta=0.25A$ and for different coupling parameter : (a) $t_{\rm TM}$=0, (b) 0.1$A$, (c) 0.2$A$, (d) 0.3$A$, (e) 0.4$A$, (f) 0.5$A$. The red, blue and green colors correspond to contributions from FM layers, top TI layer and bottom TI layer. The black lines correspond to the bulk TI bands.}
\label{bands}
\end{figure}

The band structure of a system with 20 TI layers and 5 FM layers is shown in Fig.~\ref{bands} for various coupling parameters. At $t_{\rm TM}=0$, one distinguishes the surface Dirac cones (overlapping blue and green), the TI bulk states (black) and the uncoupled magnetic states (red). Upon turning on the interlayer coupling, $t_{\rm TM}$, the Dirac cone of the top surface (blue) is pushed downward and acquires a gap due to proximity effect. The magnetic states progressively hybridize with the surface states away from the $\Gamma$-point and acquire a chiral spin texture. Notice that the Dirac cone at the bottom surface remains unaffected.

\begin{figure}[h]
\centering
\includegraphics[width=0.45\textwidth]{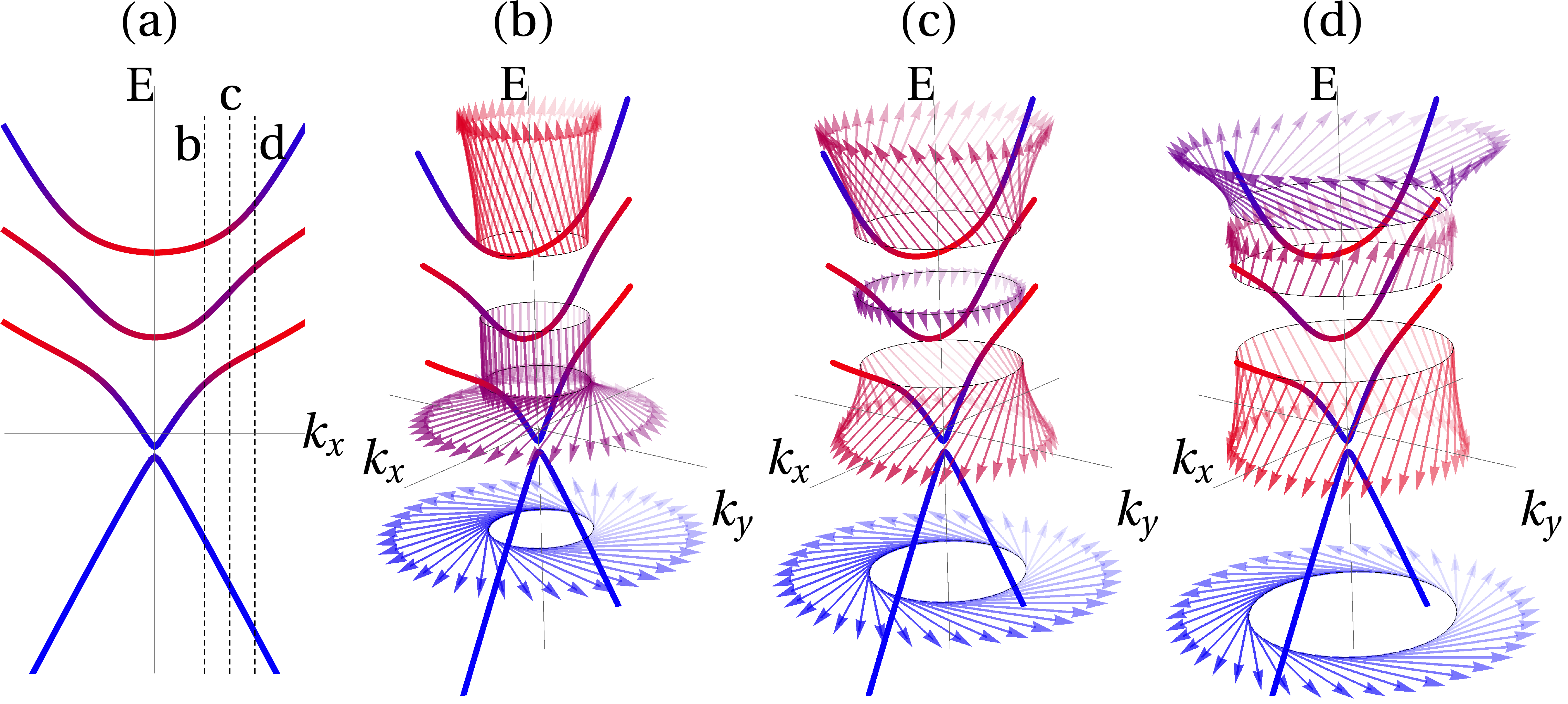}
\caption{(Color online) (a) Schematic of the band structure of a typical FM-TI heterostructure. The blue and red colors correspond to contributions from TI and FM layers, respectively. (b, c, d) Corresponding non-collinear spin texture computed at different positions in the band structure, as denoted by the vertical lines in (a).}
\label{fig1}
\end{figure}

The modification of spin texture due to coupling in a typical FM-TI is demonstrated schematically in Fig.~\ref{fig1}, which is instrumental to understand the modified spin texture in a multilayer system in the next section. In order to provide a comprehensive picture of the complex spin-momentum locking taking place in this structure, we choose three different momenta denoted by vertical dashed lines in Fig.~\ref{fig1}(a) and the corresponding spin texture is reported on Figs. \ref{fig1}(b, c, d), respectively. In a nutshell, one can notice that bands with a dominant TI character (blue) display a Dirac spin-momentum locking, ${\bf S}\sim{\bf z}\times{\bf k}$, while bands with a dominant FM character (red) display a spin angular momentum ${\bf S}\sim{\bf z}$. In general, the spin texture lies in-between these two cases. Notice that the induced chirality of the FM bands changes sign with the magnetization of the bands (denoted by red arrows).

\subsection{Non-equilibrium transport formalism}

The SOT is calculated within the linear response framework \cite{Freimuth2014, Li2015, Wimmer2016}. We start by defining the retarded (advanced) Green's function at energy $E$,
\begin{eqnarray}
\hat G^{R(A)}_{E,{\bf k}}=[(E \pm i\eta)\hat{\mathbb{I}}_n - \hat H({\bf k})]^{-1}
\end{eqnarray}
where $\eta$ is the impurity broadening and $\mathbb{I}_n$ is an $n \times n$ identity matrix, $n$ being the dimension of $\hat H$.\par

In this work, disorder is modeled by a constant broadening $\eta$ which is found to be a fair approximation for calculating SOT in Co-Pt heterostructure \cite{Freimuth2014}. We emphasize that the proper treatment of spin-independent disorder should account for vertex corrections. Such corrections are well known to cancel spin Hall conductivity \cite{Inoue2004} as well as damping-like torque \cite{Qaiumzadeh2015} in two-dimensional Rashba gas. This accidental cancellation of intrinsic SHE is an artifact of the parabolic band dispersion that results in a Berry's curvature that is independent on momentum \cite{Murakami2004b, Krotkov2006,Sinova2015}. In more complex systems, vertex corrections modify quantitatively the transport properties but do not affect their qualitative behavior \cite{Dyrdal2015, Kim2015}. Therefore, in order to keep the computation time reasonable, vertex corrections are not taken into account in our calculations. \par

To compute the SOT, we first calculate the non-equilibrium spin density caused by an applied electric field ${\bm{\mathcal E}}$. Without any loss of generality we assume the the electric field to be along $\hat{x}$ direction and incorporated as an interaction term $ev_x\mathcal{E}$. The non-equilibrium spin density per unit electric field is described by velocity-spin correlation functions \cite{Sinitsyn2007, Li2015} and possesses two contributions
\begin{eqnarray}
{\bf S} = {\bf S}_{\rm sea}+ {\bf S}_{\rm sur}
\label{s}
\end{eqnarray}
where ${\bf S}_{\rm sea}$ and ${\bf S}_{\rm sur}$ correspond contribution coming from Fermi sea and Fermi surface and are defined by,
\begin{eqnarray}
{\bf S_{\rm sur}} &=& \frac{e\hbar}{2 \pi} \int \frac{d^2 {\bf k}}{(2\pi)^2} ~ {\rm Re}({\rm Tr} [ \hat{\bf s} \hat G^R_{E,{\bf k}} \hat v_x^{\bf k} (\hat G^A_{E,{\bf k}}-\hat G^R_{E,{\bf k}}) ])_{E_F} \label{w1} \nonumber \\
\\
{\bf S_{\rm sea}} &=& \frac{e\hbar}{2 \pi} \int_{-\infty}^{E_F} dE \int \frac{d^2 {\bf k}}{(2\pi)^2} ~ {\rm Re}\left( {\rm Tr} [ \hat{\bf s} \hat G^R_{E,{\bf k}} \hat v_x^{\bf k} \partial_E \hat G^R_{E,{\bf k}} \right. \nonumber \\
&&\hspace{3.5cm} \left. - \hat{\bf s} \partial_E \hat G^R_{E,{\bf k}} \hat v_x^{\bf k} \hat G^R_{E,{\bf k}}] \right) \label{w2} \\
\hat{\bf s}_n &=& \begin{pmatrix} \ddots & & \\ & \hat{\bm \sigma}_n & \\ & & \ddots \end{pmatrix},\; \hat v_x^{\bf k} =\partial_{\hbar k_x} \hat H({\bf k}).
\label{vs}
\end{eqnarray} 
where $\hat{\bf \sigma}_n$ is the spin operator for the $n$th layer ($\hat{\bm\sigma}$ for FM layers and $\hat{\bm\sigma}\otimes\hat{\mathbb{I}}_2$ for TI layers). The integration over ${\bf k}$ goes over the first Brillouin zone $[\pm \pi/a_0, \pm \pi/a_0]$. ${\rm Re}$ takes the real part and ${\rm Tr}$ is the trace on both spin and orbital spaces. In the case of our slab geometry, we find that the Fermi sea contribution, ${\bf S}_{\rm sea}$, is negligible compared to the Fermi surface contribution ${\bf S}_{\rm sur}$. The spin-orbit torque is defined ${\bf T}=(2\Delta/\hbar) {\bf z}\times{\bf S}$, and therefore $S_x$ and $S_y$ produce the damping, $T_D$, and field-like torque, $T_F$, respectively.

One can similarly calculate the longitudinal charge conductivity using the velocity-velocity correlation function,
\begin{eqnarray}
\sigma_{xx} &=& \sigma_{\rm sea} + \sigma_{\rm sur} \label{vx} \\
\sigma_{\rm sur} &=& \frac{e^2\hbar}{2 \pi} \int \frac{d^2 {\bf k}}{(2\pi)^2} ~ {\rm Re}({\rm Tr} [ \hat{v}_x^{\bf k} \hat G^R_{E,{\bf k}} \hat{v}_x^{\bf k} (\hat G^A_{E,{\bf k}}-\hat G^R_{E,{\bf k}}) ])_{E_F} \label{wv1} \nonumber \\
\\
\sigma_{\rm sea} &=& \frac{e^2\hbar}{2 \pi} \int_{-\infty}^{E_F} dE \int \frac{d^2 {\bf k}}{(2\pi)^2} ~ {\rm Re}\left( {\rm Tr} [ \hat{v}_x^{\bf k} \hat G^R_{E,{\bf k}} \hat{v}_x^{\bf k} \partial_E \hat G^R_{E,{\bf k}} \right. \nonumber \\
&&\hspace{3.5cm} \left. - \hat{v}_x^{\bf k}\partial_E \hat G^R_{E,{\bf k}} \hat v_x^{\bf k} \hat G^R_{E,{\bf k}}] \right) \label{wv2}
\end{eqnarray}

%vanishes for all the energy range considered, and hence we only compute the Fermi surface contribution, $S_{\rm sur}^{ij}$. Without lack of generality, we consider the applied electric field along ${\bf x}$ and compute
%\begin{eqnarray}
%{\bf S} &=& \frac{e\hbar}{2 \pi} \int \frac{d^2{\bf k}}{(2\pi)^2} ~ {\rm Re}\left({\rm Tr} [ \hat{\bf s} \hat G^R_{E} \hat v_x (\hat G^A_{E}-\hat G^R_{E}) ]\right)_{E_F} \label{sx} \nonumber \\
%\\
%{\sigma_{xx}} &=& \frac{e^2 \hbar \mathcal{E}}{2 \pi} \int \frac{d^2{\bf k}}{(2\pi)^2} ~ {\rm Re}\left({\rm Tr} [ \hat v_x \hat G^R_{E} \hat v_x (\hat G^A_{E}-\hat G^R_{E}) ]\right)_{E_F} \label{vx} \nonumber \\
%\\
%{\bf S} &=& \frac{e \hbar}{2 \pi a_0^2}E_x {\rm Re}[{\rm Tr}\langle \hat{\bf s} \hat G^R_E \hat v_x (\hat G^A_E-\hat G^R_E) \rangle ]_{E_F}, \label{sx}\\
%\sigma_{xx} &=& \frac{e^2 \hbar}{2 \pi a_0^2} {\rm Re}[{\rm Tr}\langle \hat v_x \hat G^R_E \hat v_x (\hat G_E^A-\hat G^R_E) \rangle ]_{E_F}, \label{vx}\\
%\hat {\bf s} &=& \begin{pmatrix} \ddots & & \\ & \hat{\bm \sigma} & \\ & & \ddots \end{pmatrix},\; \hat v_x^{{\bf k}} =\partial_{\hbar k_x} \hat H({\bf k}).
%\end{eqnarray}

Finally, it is also useful to determine the spin Hall current flowing in the {\em bulk} of the TI as SHE is considered as a source for large damping-like SOT. The bulk TI Hamiltonian $\hat H_{B}$ and spin current operator $\hat{\bf j}_z$ are given by 

\begin{eqnarray}
&&\hat H_{B}({\bf k}) = \hat H_0+\hat H_T e^{-ik_z a_0}+\hat H_T^\dagger e^{+ik_z a_0},\\
&&\hat{\bf j}_z =(\hbar/4) \{\hat v_z^B, \hat{\bf s}^B \},
\end{eqnarray}
where $\hat v_z^B =\partial_{\hbar k_z} \hat H_B$, and $\hat {\bf s}^B = \hat{\bm \sigma }\otimes\hat{\mathbb{I}}_2$. The bulk spin Hall conductivity, $\sigma_z^i$, for a spin current polarized along $i$ and flowing along $z$ can be calculated by replacing $\hat {\bf s}$ with $\hat{\bf j}_z$ in Eqs. \eqref{w1}-\eqref{w2} and performing a three-dimensional integration over the Brillouin zone. Unlike the slab geometry we find that the Fermi sea term, Eq. \eqref{w2}, does not vanish for the bulk $\sigma_z^y$ and rather provides the main contribution to the quantized Hall conductivity within the bulk gap region. $\sigma_z^x$ and $\sigma_z^z$, on the other hand, vanish for the whole energy range.
 
%%%%%%%%%%%%%%%%%%%%%%%%%%%%%%%%%%%%%%%%%%%
\section{Results and Discussion}

\begin{figure}[h]
\centering
\includegraphics[width=0.4\textwidth]{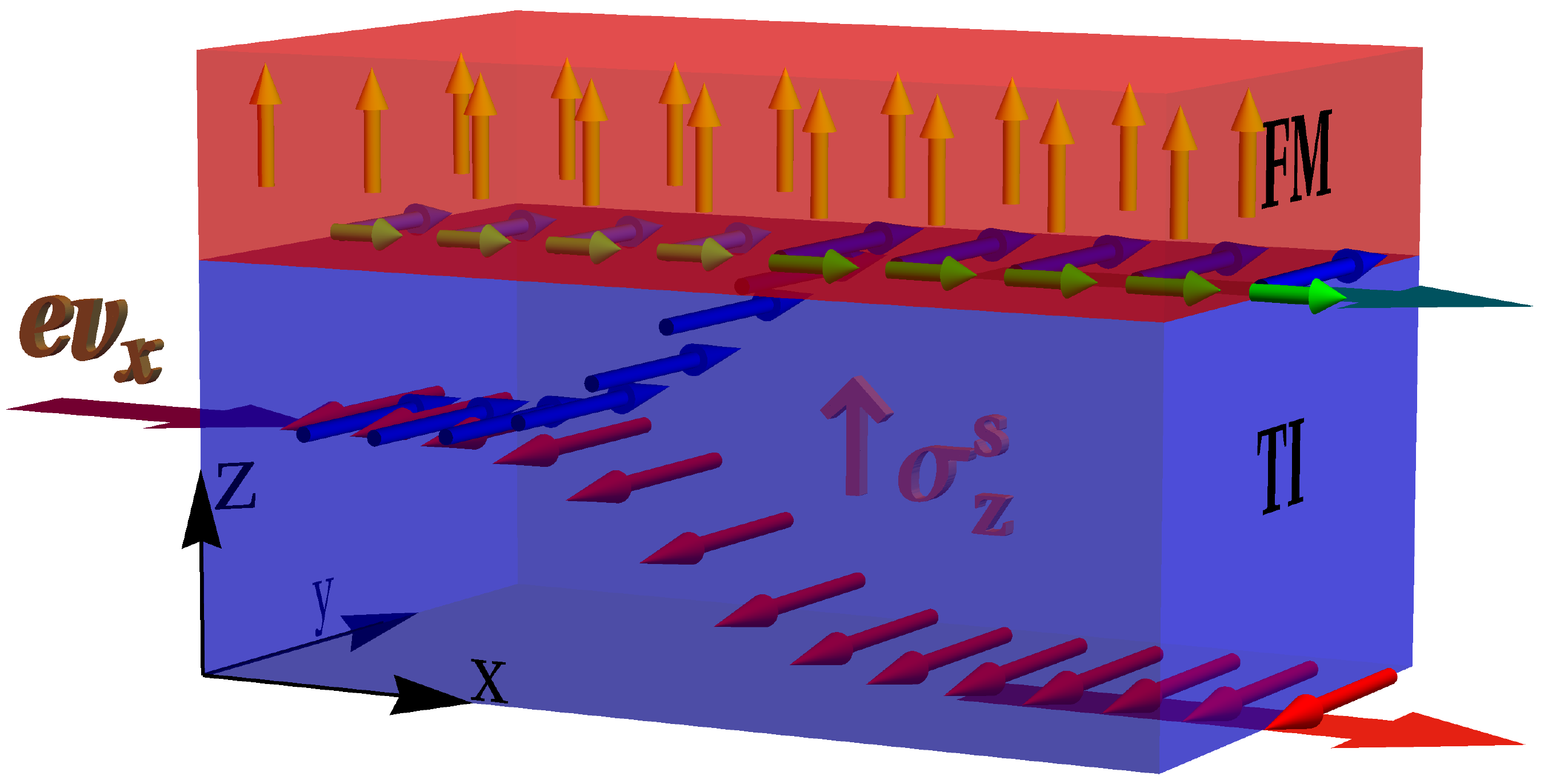}
\caption{(Color online) Schematic of the origin of different non-equilibrium spin components in an FM-TI heterostructure. In the bulk of the TI away from the interface, SHE spatially separates the flowing electrons with spin oriented along $\pm{\bf y}$ (large blue and red arrows). Electrons polarized along $+{\bf y}$ penetrates into the FM layer, generating an effective spin density oriented along $+{\bf x}$ (small green arrows). In addition, ISGE at the interface directly generates a spin density oriented along $+{\bf y}$ (small blue arrows). Both spin densities exert a torque on the magnetization of the FM layer (yellow arrows).}
\label{SOT}
\end{figure}

In magnetic multilayers SOT can be caused by two mechanisms, (i) SHE arising from the bulk of the heavy metal and (ii) ISGE induced by the interfacial (Rashba, Dirac) SOC, as illustrated in Fig.~\ref{SOT}. These two mechanisms are {\em a priori} distinct from each other. SHE generates a {\em spin current} in the bulk of the heavy metal that is injected into the adjacent FM layer; the resulting non-equilibrium spin density penetrates inside the FM layer, precesses about the magnetization vector and generates a spatially oscillating spin density. If spin dephasing is strong enough (in the case of a strong FM for instance), $S_x$ component (i.e., the damping-like torque ${\bf T}_D$) dominates \cite{Haney2013}. \par

In contrast, ISGE generates a {\em spin density} at the interface between the heavy metal and the ferromagnet. While interfacial SOC alone generates an $S_y$ component (i.e., ${\bf T}_F$) through the extrinsic Rashba-Edelstein effect\cite{Edelstein1990}, the coexistence of SOC and magnetic exchange generates an $S_x$ component (i.e., ${\bf T}_D$) through the intrinsic magnetoelectric effect\cite{Li2015,Garate2010,Ndiaye2017}. This effect is related to the Berry curvature in mixed spin-momentum space \cite{Kurebayashi2014,Manchon2014}. In TIs, the surface properties are caused by their bulk topology \cite{Isaev2011}, so one can expect a tight connection between SHE and ISGE. We now aim at understanding the interplay between SOT arising from transport in the bulk TI and SOT arising from interfacial SOC.

\subsection{Spin Hall conductivity in the bulk}

\begin{figure} [h]
\centering
\includegraphics[width=0.48\textwidth]{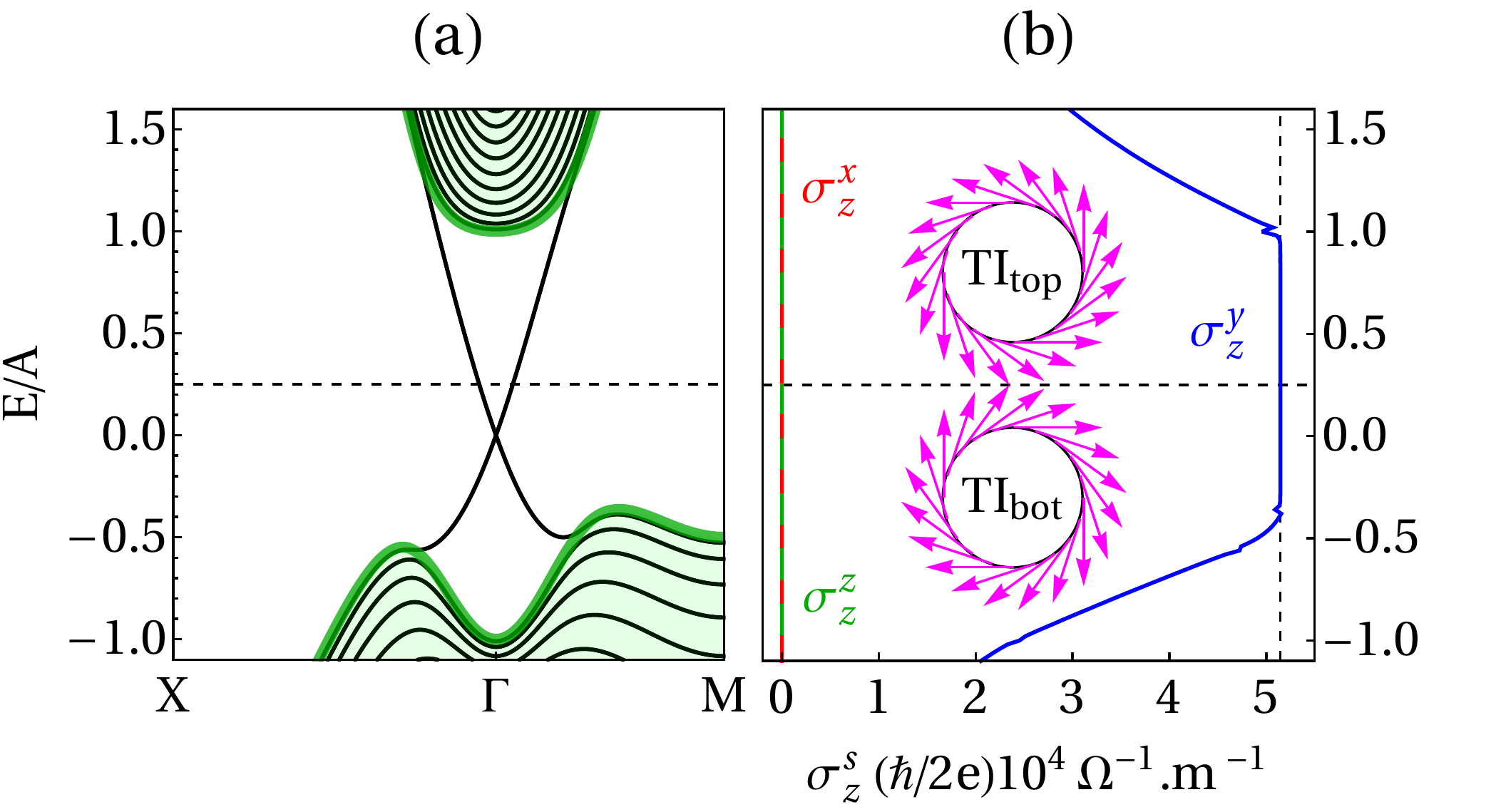}
\caption{(Color online) (a) Band structure of bulk TI (thick green lines) and a 20-layer TI slab (black lines). The corresponding spin texture of the top and bottom layers of the slab at $E=0.25A$ is represented on the insert of (b). (b) Spin Hall conductivities as a function of the energy computed in the bulk of the same TI. The spin Hall conductivity is constant as long as the energy lies in the bulk gap and decreases when bulk states become conductive.}
\label{bulkSHE}
\end{figure}

To clarify these aspects, we first calculate the bulk spin Hall conductivity and the surface spin texture of a TI slab with 20 layers in the absence of magnetic overlayer. The band structures of the bulk TI (thick green) together with the one of the equivalent 20-layer slab (black) are displayed on Fig.~\ref{bulkSHE}(a); the spin texture at $E=0.25A$ for both top and bottom surfaces is reported as an inset on Fig.~\ref{bulkSHE}(b). The corresponding {\em bulk} spin Hall conductivity is displayed in the main panel of Fig.~\ref{bulkSHE}(b). For $-0.5<E/A<1$, the TI is bulk insulating and only conducts through its surface states. Hence, the spin Hall conductivity is constant, $\sigma_z^y \approx 5\times10^4$~$(\hbar/2e)$~$\Omega^{-1}\cdot$m$^{-1}$, i.e., an electric field applied along ${\bf x}$ creates a spin current polarized along ${\bf y}$ and propagating along ${\bf z}$. For $E/A>1$, the TI is conductive, which results in a progressive decrease of the spin Hall conductivity $\sigma_z^y$. For the sake of comparison, the spin Hall conductivity of 5$d$ transition metals such as Pt, Ta or W has been calculated to be in the range of $10^3$ to $10^4$ $(\hbar/2e)$~$\Omega^{-1}\cdot$m$^{-1}$ (Ref. \onlinecite{Tanaka2008}), while a maximum of $10^5$ $(\hbar/2e)$~$\Omega^{-1}\cdot$m$^{-1}$ was computed for Bi$_{0.83}$Sb$_{0.17}$ \cite{Sahin2015}. The value of the intrinsic spin Hall conductivity reported in Fig.~\ref{bulkSHE} is therefore quite large and should generate large damping-like SOT.

\subsection{Surface versus Bulk Transport}

\begin{figure}[h]
\centering
\includegraphics[width=0.45\textwidth]{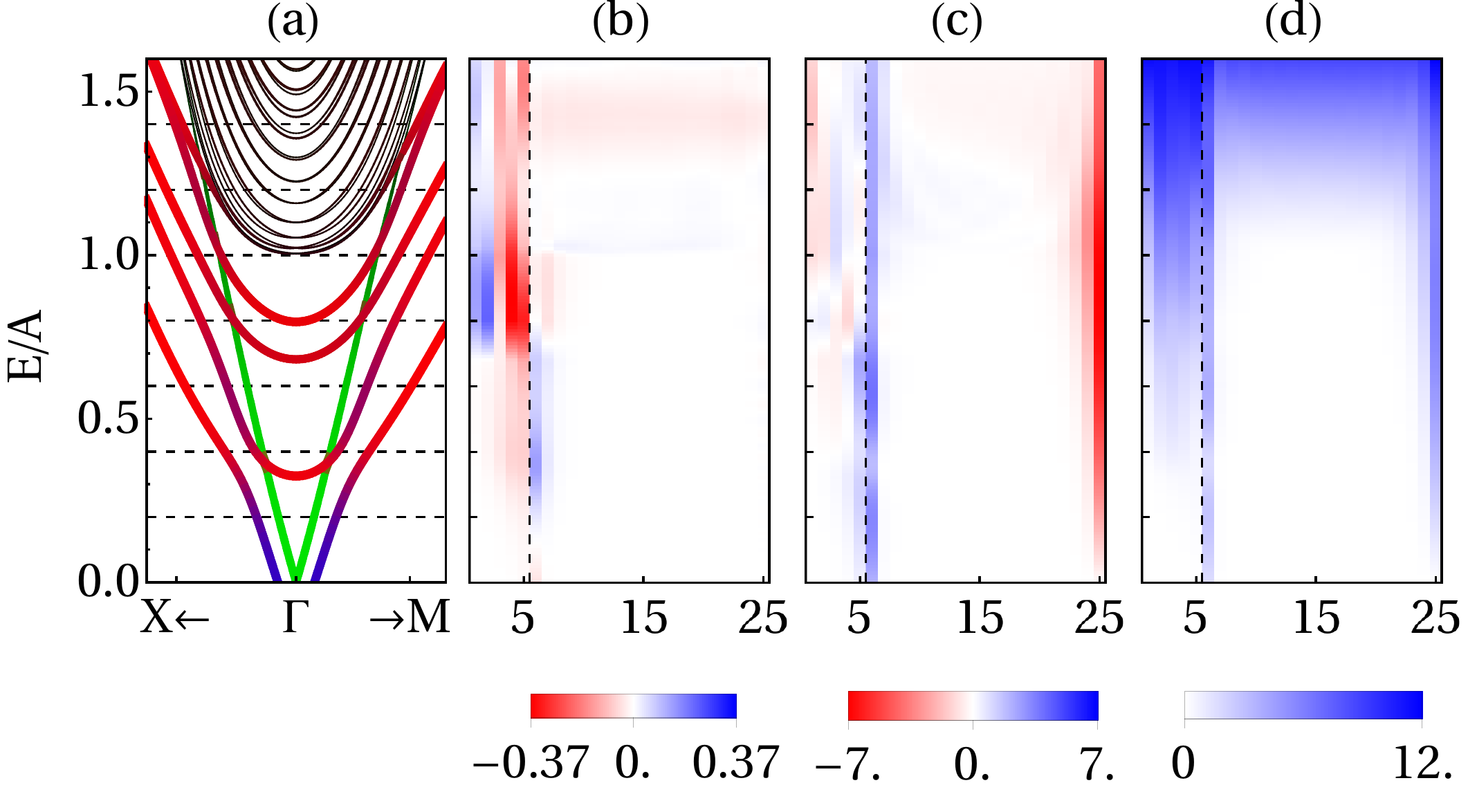}
\caption{(Color online) (a) Layer-resolved band structure of the 5 FM + 20 TI slab, energy-resolved non-equilibrium spin density, (b) $S_x$ and (c) $S_y$, and (d) longitudinal charge conductivity $\sigma_{xx}$ across the FM-TI heterostructure, with $t_{\rm TM}=0.5A$ and $\Delta=0.25A$. In (b,c,d) the FM-TI interface is marked by a vertical dashed line. The color scale gives $S_{x,y}$ in units of $10^{10}$ V$^{-1}\cdot$m$^{-1}$ and $\sigma_{xx}$ in $10^{-4}~\Omega^{-1}$.}
\label{elayer}
\end{figure}

Figure~\ref{elayer} shows the spatial profile of (b) $S_x$, (c) $S_y$ and (d) $\sigma_{xx}$ for a system with 5 FM and 20 TI layers, while tuning the transport energy through the band structure [Fig.~\ref{elayer}(a)]. Within the bulk gap region ($-0.5<E/A<1$), $S_y$ and $\sigma_{xx}$ are mostly coming from TI surfaces. The magnetic bands are crossing the surface states at $E\sim 0.3A$, which is denoted by a progressive rise in $S_x$, $S_y$ and $\sigma_{xx}$ within the magnetic layer and a drop in $S_y$ in the top TI surface (close to FM layer). Interestingly, the rise of $S_x$ and collapse of $S_y$ at the top TI surface are correlated with the {\em onset of bulk conduction}, when $E>A$.

\begin{figure}[h]
\centering
\includegraphics[width=0.5\textwidth]{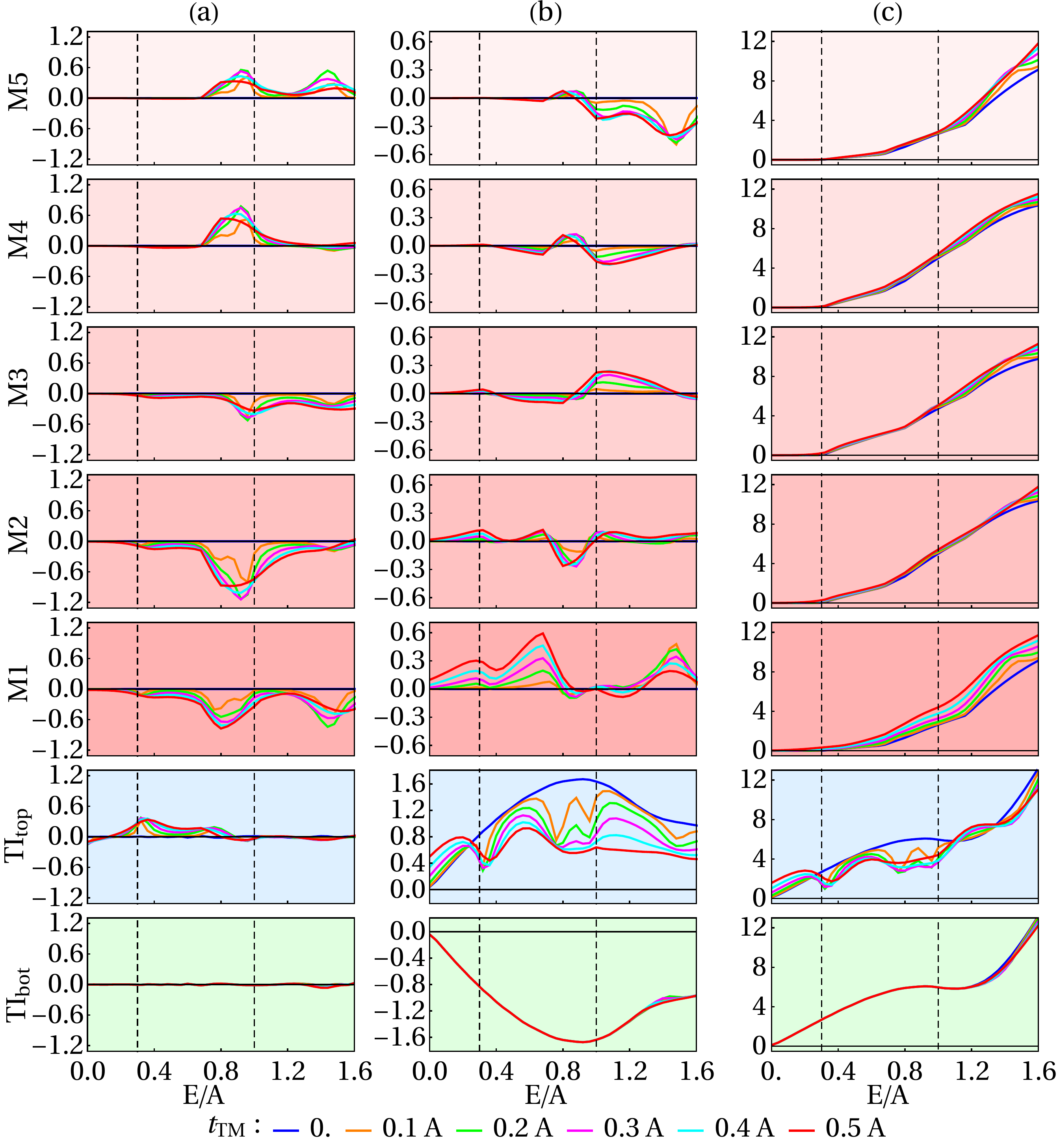}
\includegraphics[width=0.35\textwidth]{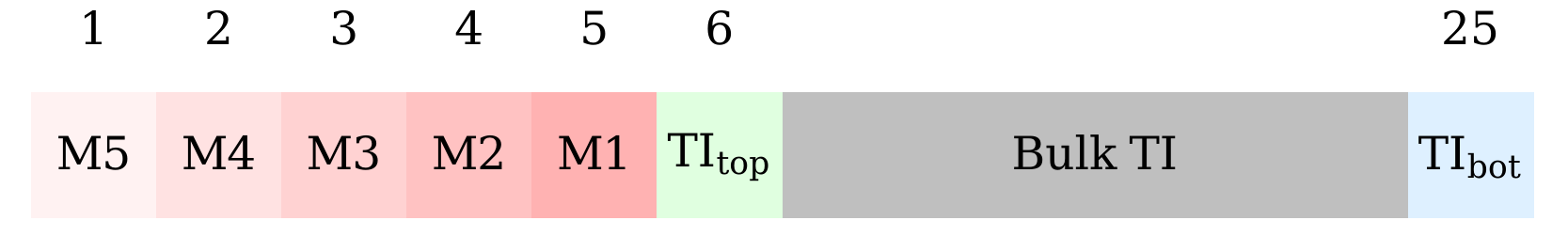}
\caption{(Color online) (a) $S_x$, (b) $S_y$ and (c) $\sigma_{xx}$ for a 5 FM + 20 TI heterostructure for different coupling constant ($t_{\rm TM}$) with exchange coupling $\Delta=0.25A$ and broadening $\eta=0.01A$. The vertical dashed lines show the beginning of decoupled magnetic bands ($E=0.3A$) and bulk TI bands ($E=1.0A$). $S_{x}$ is given in units of $10^{9}$ V$^{-1}\cdot$m$^{-1}$, $S_{y}$ in $10^{10}$ V$^{-1}\cdot$m$^{-1}$ and $\sigma_{xx}$ in $10^{-4}~\Omega^{-1}$. The bottom panel shows the color code for different layers.}
\label{sxym25}
\end{figure}

To better understand this behavior, we investigate in more details the layer-resolved spin density and conductivity as a function of the energy in selected layers. Figure~\ref{sxym25} displays the transport energy dependence of $S_x$, $S_y$ and $\sigma_{xx}$, at the top and bottom TI surfaces, TI$_{\rm top}$ and TI$_{\rm bot}$, as well as in the different FM layers, M1, M2, M3, M4 and M5, for different coupling strengths ($t_{\rm TM}$). Based on the band structure shown in Fig.~\ref{elayer}(a) and on the conductivity map of Fig.~\ref{elayer}(d), we define three transport regimes. When $0<E/A<0.3$, the transport solely occurs through the surface states of the TI and the FM behaves like a magnetic insulator. This situation is comparable to the magnetic Dirac gas studied in Refs. \onlinecite{Garate2010,Sakai2014,Ndiaye2017}. Then, for $0.3<E/A<1$, the FM layer becomes progressively conductive while the TI remains bulk insulating. Finally, when $E/A>1$ conduction occurs throughout the entire heterostructure, and the higher the transport energy the more bulk transport dominates over surface transport.\par

Let us first consider the uncoupled situation, i.e., the TI and FM layers are disconnected ($t_{\rm TM}=0$, blue line in Fig.~\ref{sxym25}). When varying the transport energy across the band structure, $S_y$ progressively increases at both TI$_{\rm top}$ and TI$_{\rm bot}$, while $S_x$ remains exactly zero. $S_y$ reaches a maximum close to the bulk conduction edge ($E/A\approx 1$) and decreases monotonously for $E/A>1$, consistently with the behavior of the bulk spin Hall conductivity displayed on Fig.~\ref{bulkSHE}. Obviously, the FM layers do not show any non-equilibrium spin density because $t_{\rm TM}=0$.\par

\begin{figure}[h!]
\centering
\includegraphics[width=0.45\textwidth]{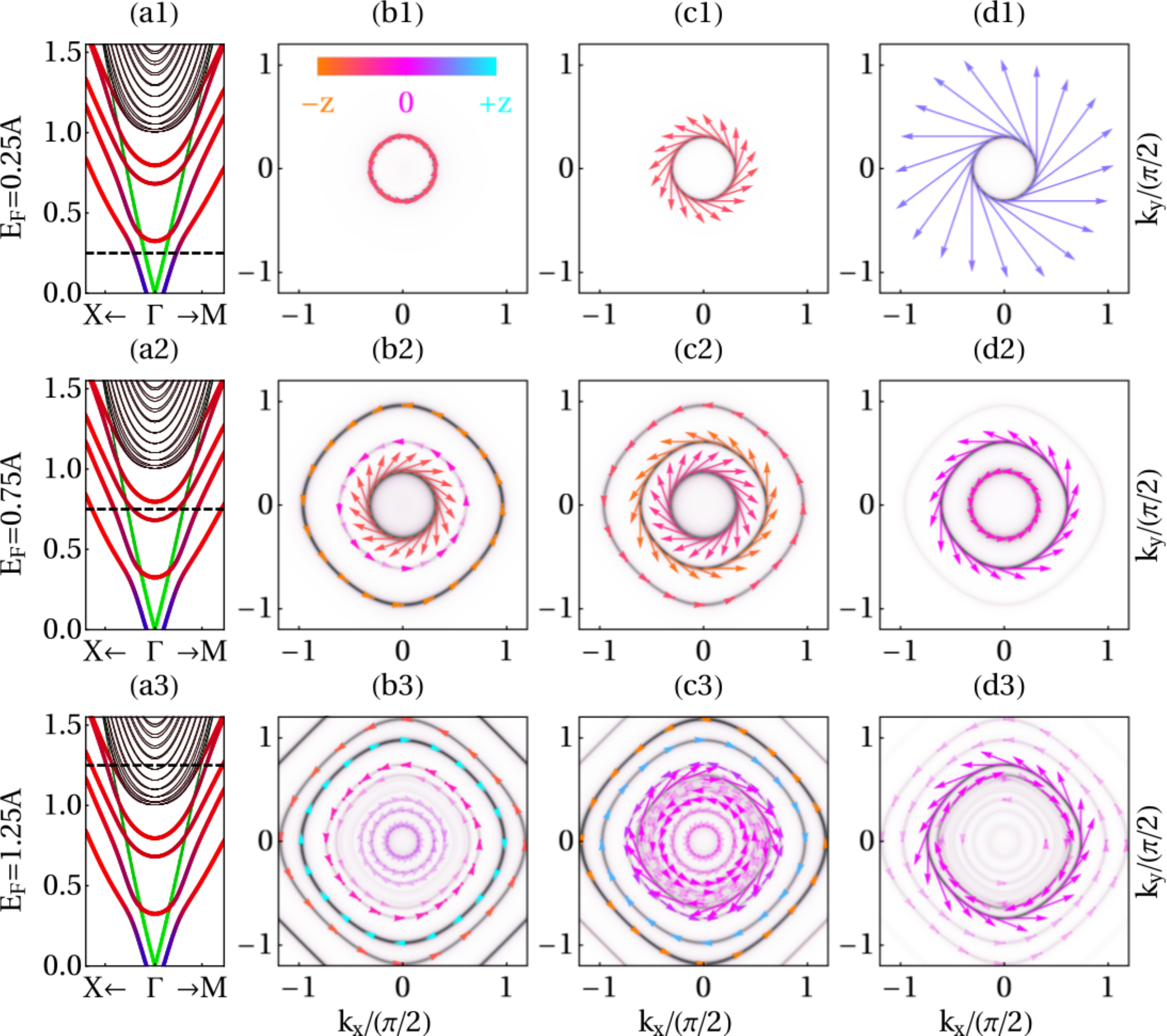}
\caption{(Color online) (a1, a2, a3) Band structure and Spin texture of (b1, b2, b3) M1 and (c1, c2, c3) M2, and (d1, d2, d3) TI$_{top}$ at (1) $E=0.25A$, (2) $E=0.75A$ and (3) $E=1.25A$ for $t_{TM}=0.5A,\;\Delta=0.25A$ and $\eta=0.01A$. The texture colors correspond the $S_z$ component of the spin density. The band colors have the same meaning as Fig.~\ref{bands}.}
\label{sptxt}
\end{figure}

When the FM-TI coupling is turned on, the FM bottom band acquires a Rashba-like SOC, while a gap opens at the Dirac cone at TI$_{\rm top}$. To apprehend the physics at stake, the spin texture in three different energy regimes has been plotted on Fig.~\ref{sptxt}, for $t_{\rm TM}=0.5$. As mentioned above, when $E/A<0.3$ the transport is dominated by the Dirac states of the TI surfaces, while the FM is insulating. Turning on the FM-TI coupling pushes the Dirac cone downwards [see Fig.~\ref{sptxt}(a1)] and thereby enhances the density of states at Fermi level resulting in an increase in both $S_y$ and $\sigma_{xx}$ at TI$_{\rm top}$ (Fig.~\ref{sxym25}). The first two magnetic layers, M1 and M2, become weakly conductive by proximity effect and acquire a small spin texture aligned on the one of TI$_{\rm top}$ [see Fig.~\ref{sptxt}(b1,c1,d1)]. As a result, $S_y$ penetrates into the FM layers by proximity and the larger the FM-TI coupling, the stronger the induced spin density. From the gap of TI$_{\rm top}$ bands, we can estimate that for $t_{\rm TM}=0.5$ the induced magnetic exchange is roughly 25\% of the FM exchange coupling ($\Delta$). Consequently, an $S_x$ component progressively appears in both TI$_{\rm top}$ and M1 via the magnetoelectric effect\cite{Kurebayashi2014,Garate2010}, due to the coexistence of magnetism and SOC. Yet, this component remains extremely small.\par

In the intermediate regime, $0.3<E/A<1$, the TI layer is still in a topologically non-trivial state (bulk insulating, conductive chiral surface states), while the FM layers become more conductive and acquire a complex spin texture whose chirality depends on the magnetization, as displayed in Fig.~\ref{sptxt}(b2,c2). On the other hand, the strength of the spin-momentum locking at TI$_{\rm top}$ decreases upon increasing the transport energy [Fig.~\ref{sptxt}(d2)]. Consequently, the competition between the different spin chiralities produces an oscillating behavior of $S_y$ as a function of the energy: the dips correspond to FM-TI band crossing. Notice that increasing the coupling results in a {\em decrease} of $S_y$ at TI$_{\rm top}$ and an {\em increase} in the FM layers (Fig.~\ref{sxym25}). Indeed, upon increasing the coupling, the FM layers acquire more SOC and because the FM is now conducting, $S_y$ penetrates deeply into the FM layers producing a $S_x$ component upon spin precession. $S_x$ also increases in the FM layers upon increasing the FM-TI coupling. 

Finally, when $E/A>1$ the transport is progressively dominated by the bulk states of the TI. The central portion of the texture is coming from bulk TI and depending on whether states are dominated by the top TI layer or the hybridization of the FM layer, it can have either positive or negative chirality. 
 As shown in Fig.~\ref{sptxt}(d3), the TI bands possess weaker spin-momentum locking, causing a fall in $S_y$ in TI$_{\rm top}$ and a deeper penetration inside the FM layers. This penetration is associated with spin precession in the FM, and therefore a significant increase in $S_x$, as displayed in Fig.~\ref{sxym25}. Because the wavefunctions are now delocalized, they expand throughout the structure, from M5 to TI$_{\rm bot}$ [Figs.~\ref{elayer}(b,c)].

In summary, while the $S_y$ component dominates in the surface-dominated regime, in agreement with all previous theories on TI \cite{Garate2010,Sakai2014,Ndiaye2017}, a crossover appears upon increasing the transport energy and in the bulk-dominated regime, the $S_x$ component is significantly enhanced inside the FM layer, displaying a large oscillation across the thickness. This is consistent with the onset of the SHE in the bulk TI. It is worth noticing that the magnitude of the $S_x$ component is about the same as $S_y$, so at this stage it is unclear whether these results can explain the experimental observations. One possible reason could be the fact that we are modeling each FM layer with only a pair of parabolic bands, whereas in practice a real FM material has many more bands crossing Fermi level. Therefore our results are expected to be up to an order of magnitude smaller than reported experimentally. Nevertheless, the physical arguments are still valid for realistic heterostructures. To this end, we now address the impact of disorder on the magnitude of the SOT components, in order to identify their physical origin.

\subsection{Intrinsic versus Extrinsic Spin-Orbit Torque}

To better understand the spin-charge conversion mechanisms at stake in the FM-TI heterostructure, we investigate the dependence of the spin density as a function of the impurity broadening $\eta$. As a matter of fact, it was shown recently that in the context of the Rashba two-dimensional electron gas or in the bulk of (Ga,Mn)As, both interband and intraband processes participate to the SOT and give rise to two classes of contributions \cite{Kurebayashi2014,Li2015,Ndiaye2017}: (i) extrinsic contributions that depend on the amount of disorder in the system and (ii) intrinsic contributions that are independent on the amount of disorder. ISGE is an extrinsic mechanism and is expected to exhibit a $1/\eta$ dependence, like the conductivity, while the magnetoelectric effect and SHE are both intrinsic contributions and should not vary as a function of $\eta$. In fact, the former is related to the Berry curvature in mixed spin-momentum space of the states present at the interface \cite{Kurebayashi2014}, while the latter is associated with the Berry curvature in mixed momentum space of the states present in the bulk \cite{Sinova2015}.

\begin{figure}[h]
\centering
\includegraphics[width=0.45\textwidth]{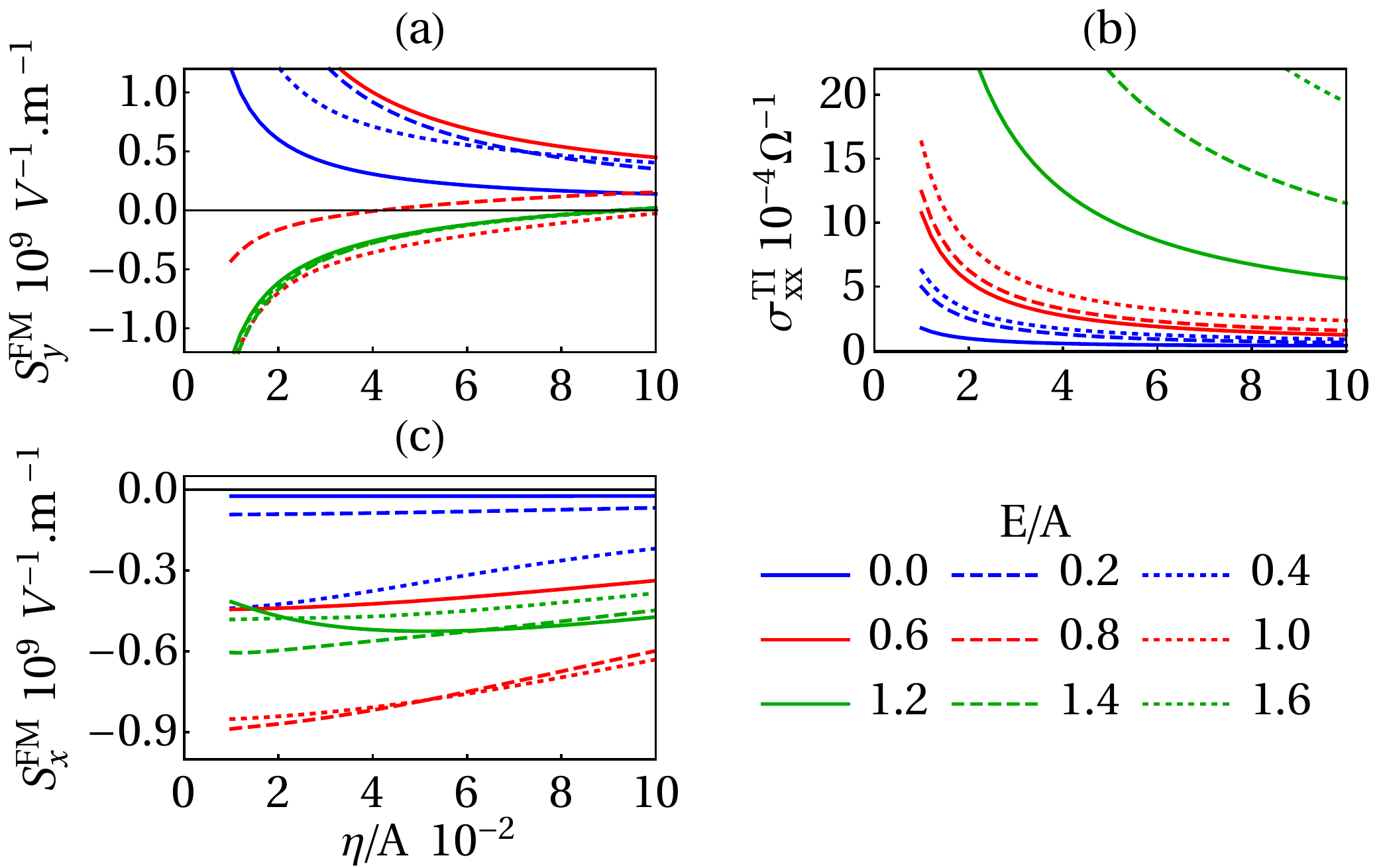}
\caption{(Color online) Non-equilibrium (a) $S_y^{\rm FM}$, (c) $S_x^{\rm FM}$ in FM and TI conductivity (b) $\sigma_{xx}^{\rm TI}$ with $t_{\rm TM}=0.5A$ and $\Delta=0.25A$ for different disorder broadening $\eta$. The blue lines correspond to the surface-dominated/magnetic insulator regime, the red lines correspond to the surface-dominated/magnetic metal regime, and the green lines correspond to the bulk-dominated/magnetic metal regime.}
\label{eta}
\end{figure}

Figure~\ref{eta} displays the dependence of the total spin density $S_{x,y}^{\rm FM}=\sum_{\rm FM} S_{x,y}$, summed over the FM layers, as well as the total conductivity of the TI, $\sigma^{\rm TI}_{xx}=\sum_{\rm TI}\sigma_{xx}$. Both $S^{\rm FM}_{y}$ and $\sigma^{\rm TI}_{xx}$ display the $1/\eta$-dependence expected for extrinsic mechanisms. It means that $S_y$ comes from the interfacial (Rashba- or Dirac-driven) ISGE, namely from the spin-momentum locked interfacial states of the heterostructure. The change of sign of $S^{\rm FM}_{y}$ for $E/A=1.2$ can be ascribed to the competition between Rashba- and Dirac-driven ISGE, as they have opposite sign. The behavior of $S^{\rm FM}_x$ is richer and depends on the transport regime. As long as the transport is purely interfacial, $S^{\rm FM}_x\sim 1+O(\eta)$, i.e., this component is independent on the disorder. Since the system is in the quantum SHE regime, the physical origin of $S_x$ is attributed to the interfacial magnetoelectric effect. In the intermediate and bulk regimes, when the FM and TI layers are conductive, $S^{\rm FM}_x$ displays a small dependence as a function of the disorder, which can be attributed to the diffusion of the spin density inside the metallic FM. Yet, $S^{\rm FM}_x$ converges to a constant value when $\eta\rightarrow0$ indicating its intrinsic origin.

\subsection{Understanding the torque efficiency\label{s:eff}}

Let us now complete this study by quantitatively comparing the magnitude of the SOT computed with our model to the one observed in experiments. Assuming $A\sim 0.27$ eV and $a_0=4.2~$\AA\ gives a group velocity $(a_0A/\hbar)\sim 1.7\times 10^5$ m/s near the Dirac cone, and a bulk band gap $1.5A = 0.4$ eV. Near $E\sim0.7A$, uncoupled ($t_{\rm TM}=0$) TI surfaces display a non-equilibrium two dimensional spin density per unit field $S_y \sim 1.5 \times 10^{10}$ V$^{-1}\cdot$m$^{-1}$ (Fig.~\ref{sxym25}). Assuming the thickness of a single TI layer to be 1 nm and an applied electric field ${\cal E}=2\times10^4$ V/m \cite{Mellnik2014}, we obtain a total spin density $0.3\times10^{24}$ m$^{-3}$ which is in good agreement with the spin density calculated from density functional theory \cite{Chang2015a}. This estimation ensures that the TI parameters chosen in the present study are realistic. The non-equilibrium spin density on the FM layers on the other hand depends on the coupling strength ($t_{\rm TM}$) as well as on the magnetic exchange. In the strong coupling limit ($t_{\rm TM}=0.5A$), within the previous parameter settings, the first FM layer (M1) can acquire a spin density $S_y$ that amounts up to 20\% of the value of the decoupled TI surface. \par

In experiments, the SOT is measured as a magnetic field $H_{\rm SOT}$ (in Oe or T) per unit of current density flowing in the heavy metal $j_e$ (in A/cm$^2$). As briefly explained in the introduction, it is now conventional to quantify the torque in terms the spin conductivity $\sigma_s=H_{\rm SOT}M_sd/E$ [in $(\hbar/2e)~\Omega^{-1}\cdot$m$^{-1}$] and the dimensionless SOT efficiency $\theta_{\rm H}=(2e/\hbar)\sigma_s/\sigma_{xx}^{\rm HM}$, where $M_s$ is the saturation magnetization, $d$ is the FM thickness, and $\sigma_{xx}^{\rm HM}$ is the conductivity of the adjacent heavy metal. It is understood that $\theta_{\rm H}$ is equal to the spin Hall angle if and only if SHE is the only spin-charge conversion mechanism present in the system - which is clearly not the case of transition metal bilayers or TI-FM heterostructures, as demonstrated in the present work.

\begin{table}
\begin{tabular}{cc|ccccc}
Overlayer & Ref. & $\sigma_{xx}$ & $\sigma_s^\|$ & $\sigma_s^\bot$& $\theta_{\rm H}^\|$ & $\theta_{\rm H}^\bot$ \\\hline\hline
 Bi$_2$Se$_3$(8)/NiFe(8) & \onlinecite{Mellnik2014} & 5.7 & 16 & 20 & 3.5 & 2.8 \\
%CoFeB(5) & \onlinecite{Wang2015} & & - & & - & \\ 
Bi$_2$Se$_3$(7.4)/CoTb(4.6)& \onlinecite{Han2017}& 9.4 & 1.5 & & 0.16 & \\
Bi$_2$Se$_3$(5)/CoFeB(7) & \onlinecite{Wang2017d} & 2.4 & 3.89 & & 1.6 & \\
Bi$_2$Se$_3$(4)/CoFeB(5) & \onlinecite{Mahendra2017} & 0.78 & 15 & & 18.83 & \\
Dirac gas & \onlinecite{Ndiaye2017}& 0.19 & 1.48 & 2.35 & 7.66 & 12.1
\end{tabular}
\caption{\label{tab:tidata} Room temperature bulk conductivity, spin conductivity and SOT efficiency measured in various Bi$_2$Se$_3$($t$)/FM($d$) systems, where $t$ and $d$ are the thicknesses in nm. The bulk conductivity is in $10^4~\Omega^{-1}\cdot$m$^{-1}$ and the spin conductivity is in $(\hbar/2e)~10^4~\Omega^{-1}\cdot$m$^{-1}$. The numbers in the last column are computed using Eqs. (\ref{eq:ti2d1}) and (\ref{eq:ti2d2}).}
\end{table}

\begin{figure}[t]
\centering
\includegraphics[width=0.48\textwidth]{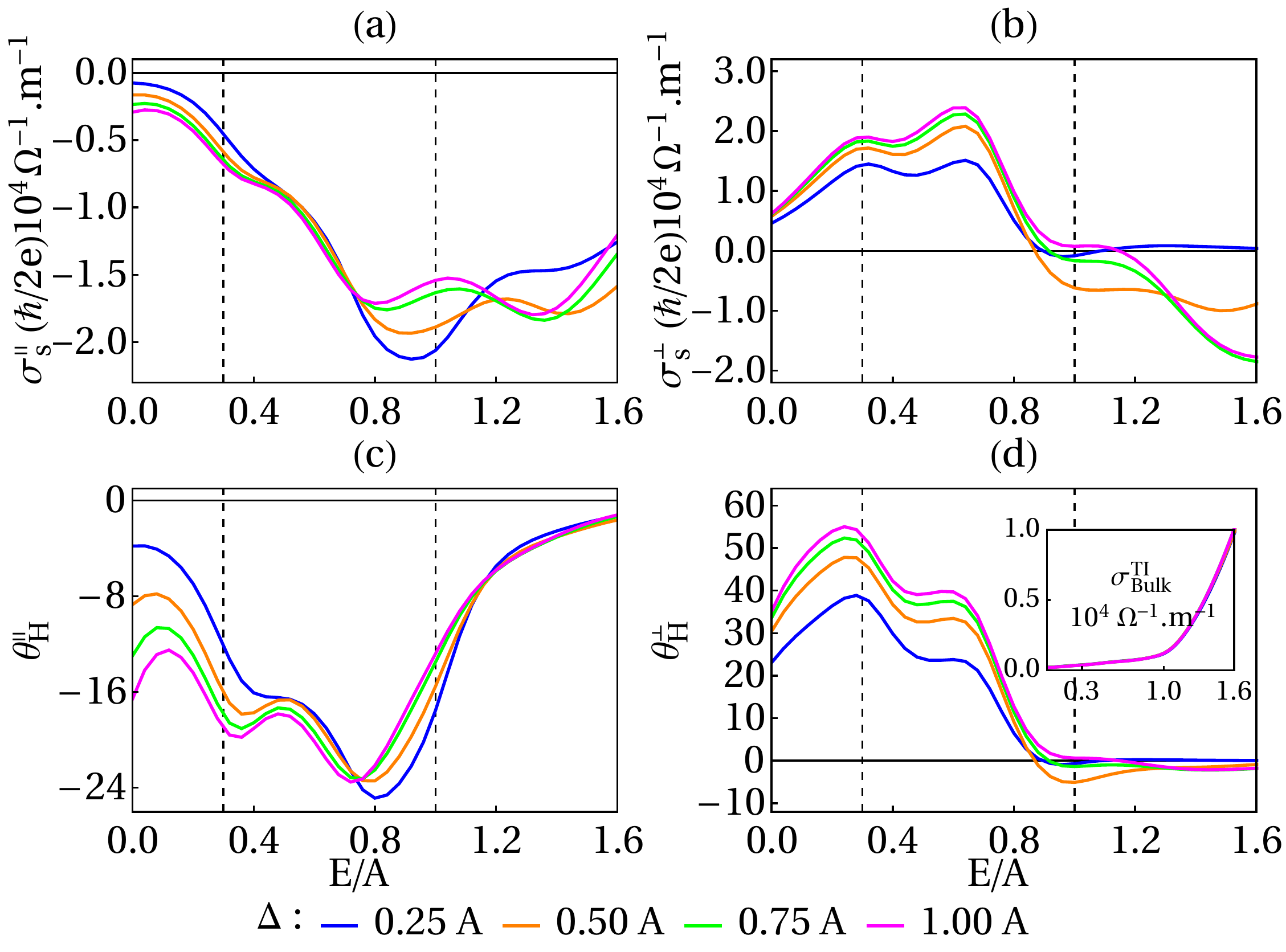}
\caption{(Color online) (a) Longitudinal and (b) transverse spin conductivity, $\sigma_s^{\|,\bot}$, corresponding to the damping-like and field-like torque components respectively, as a function of the transport energy, for a system with 5 FM + 20 TI layers with $t_{TM} = 0.5$ and $\eta = 0.1A$. Their associated effective spin Hall angle $\theta^{\|,\bot}_{\rm H}$ are reported on (c) and (d), respectively. Inset of (d) shows $\sigma_{\rm Bulk}^{\rm TI}$.}
\label{et-total}
\end{figure}

We selected four experimental works characterizing SOT on Bi$_2$Se$_3$-based heterostructures, and whose results are reported on Table \ref{tab:tidata}. Overall, the spin conductivity ranges from $10^4$ to $10^5~\Omega^{-1}\cdot$m$^{-1}$, while the SOT efficiency spans over two orders of magnitude depending on the estimated bulk TI conductivity. These data need to be taken with sane care, considering the difficulty in accurately estimating the various materials parameters (due to large interfacial roughness, magnetic dead layers \cite{Wang2015}, inhomogeneous TI conductivity etc.). In our tight-binding model, the torque density is ${\bf T}=(2\Delta/\hbar){\bf z}\times{\bf S}$, and the corresponding spin conductivity is defined $\sigma_s^{\|,\bot}=\pm(2e/\hbar)\Delta\sum_{\rm FM} S_{x,y}$. Here $\sum_{\rm FM}$ denotes the summation over the FM layers and the superscript $\|,\bot$ denotes the damping-like and field-like SOT components. The effective bulk conductivity of the TI layer is then $\sigma_{\rm Bulk}^{\rm TI}=\sum_{\rm TI}\sigma_{xx}/W$, where $W$ is the thickness of the TI and the summation is only performed on the TI layers. Figure~\ref{et-total} displays these four observables as a function of the transport energy for different magnetic exchange energies. The impurity broadening is taken as $\eta=0.1A$ ($\approx$ 27 meV), such that the TI conductivity corresponds to the one observed experimentally, close to the bulk conduction regime [see insert of Fig.~\ref{et-total}(d)].\par

We observe that the spin conductivity monotonously increases with the transport energy, reaching a maximum around $E/A\approx0.6-0.8$, which corresponds to the surface-dominated transport. Beyond this point, the spin conductivity decreases [Figs. \ref{et-total}(a) and (b)]. In other words, the maximum SOT magnitude is attained {\em before} reaching the bulk transport regime despite the large bulk spin conductivity of our system [see Fig.~\ref{bulkSHE}]. These results indicate that SHE is not the main mechanism for SOT in TIs (as the largest torque magnitude is obtained in the surface-dominated regime), but rather the interfacial ISGE. As a matter of fact in the bulk-dominated transport regime, $E/A>1$, the SOT efficiency decreases dramatically due to the increasingly large conductivity of the bulk TI states. Within our set of parameters, the computed $\sigma_s^{\bot}$ and $\sigma_s^{\|}$ are comparable to the ones observed in Refs. \onlinecite{Wang2017d} and \onlinecite{Han2017}, and one order of magnitude smaller than Refs. \onlinecite{Mellnik2014} and \onlinecite{Mahendra2017}, see Table \ref{tab:tidata} (probably due to the details of the band structure and scattering processes). Interestingly, the damping-like SOT efficiency reaches up to $\sim$ 24 ($\equiv 2,400\%$) close to the onset of the bulk-dominated transport, a magnitude comparable to the one reported in Ref. \onlinecite{Mahendra2017}. Notice though that this quantity is very sensitive to the overall bulk conductivity $\sigma_{\rm Bulk}^{\rm TI}$, which can be easily tuned by changing the disorder broadening $\eta$. While we do not intend to quantitatively fit the experimental data, these calculations compare favorably with those reported in Table \ref{tab:tidata} and clearly suggest that SHE from bulk states are inefficient to generate large SOTs in TI-FM heterostructures. 

%Wang2017 in Fig. 3: Bi$_2$Se$_3$(X)/CoFeB(7), for X=5QL, $\sigma_s^\|\approx3.89$ while for X=20QL, $\sigma_s^\|\approx3.5$.

To complete this discussion, we also report the results obtained within the Dirac model that assumes a induced magnetic exchange at the surface. This model gives\cite{Ndiaye2017}
\begin{eqnarray}\label{eq:ti2d1}
&&\sigma_s^\bot=-e\tau\Delta_{\rm ind}\varepsilon_{\rm F}/(\hbar^2v_{\rm F}\pi),\\
&&\sigma_s^\|=2e\Delta_{\rm ind}^2/(\hbar\varepsilon_{\rm F} v_{\rm F}\pi),\label{eq:ti2d2}
\end{eqnarray}
where $\Delta_{\rm ind}$ is the induced magnetic exchange (about 25 $\%$ of the magnetic exchange for $t_{\rm TM}=0.5A$), $\varepsilon_{\rm F}$ is the Fermi energy, $\tau$ is the scattering time and $v_{\rm F}$ is the Fermi velocity. The surface conductivity of the Dirac gas reads $\sigma_{xx}^{\rm 2D}=(e^2/h)(\tau\varepsilon_{\rm F}/\hbar)$, giving an effective three-dimensional conductivity $\sigma_{xx}=\sigma_{xx}^{\rm 2D}/W$ (where $W\approx20$ nm). For this estimation, we took the same parameters as for our FM-TI heterostructure, $v_{\rm F}=1.7\times10^5$ m/s, $\tau=\hbar/2\eta\approx10^{-14}$ s, $\varepsilon_{\rm F}=0.2A=54$ meV, and $\Delta_{\rm ind}=17$ meV. We obtain a spin conductivity that is comparable to ones observed experimentally, i.e., in the range $(\hbar/2e)~10^5~\Omega^{-1}\cdot$m$^{-1}$ and a small longitudinal conductivity, $\sim10^4~\Omega^{-1}\cdot$m$^{-1}$, comparable to Ref. \onlinecite{Mahendra2017}. In this experiment, Bi$_2$Se$_3$ is not grown epitaxially but by sputtering, which results in a reduced bulk conductivity and enhanced surface transport. All these considerations rule out SHE as a source of giant SOT in TI-FM heterostructures.

\section{Conclusion} 

In summary, we have developed a tight-binding model for FM-TI heterostructure that accounts for surface and bulk transport on equal footing. In addition, our minimal hybridization scheme reproduces properly the induced magnetic exchange, induced Rashba spin-orbit coupling and energy shift of the band structure observed using {\em ab initio} calculations\cite{Zhang2016}. Our study unveils the momentum-dependent spin texture across the band structure when varying the hybridization strength. This approach enables us to compute the SOT emerging from the coexistence of SHE, ISGE, magnetoelectric effect, as well as spin precession inside the ferromagnet. Our analysis shows that SOT increases steadily when increasing the transport energy and reaches a maximum {\em before} the bulk states start contributing to the transport. This result indicates that the SHE from bulk states is unlikely to explain the large damping-like SOTs observed experimentally. In contrast, large damping-like torque is achieved through the interfacial magnetoelectric effect promoted by the Berry curvature of the interfacial states and is therefore very sensitive to the nature of interfacial orbital hybridization.\par

\section{Acknowledgments}
This work was supported by the King Abdullah University of Science and Technology (KAUST). The authors would like to acknowledge support from KAUST Supercomputing facility.
%%%%%%%%%%%%%%%%%%%%%%%%%%%%%%%%%%%%%%%%%%%%
\bibliographystyle{apsrev4-1}
\bibliography{FMTI}

\begin{thebibliography}{82}%
\makeatletter
\providecommand \@ifxundefined [1]{%
 \@ifx{#1\undefined}
}%
\providecommand \@ifnum [1]{%
 \ifnum #1\expandafter \@firstoftwo
 \else \expandafter \@secondoftwo
 \fi
}%
\providecommand \@ifx [1]{%
 \ifx #1\expandafter \@firstoftwo
 \else \expandafter \@secondoftwo
 \fi
}%
\providecommand \natexlab [1]{#1}%
\providecommand \enquote  [1]{``#1''}%
\providecommand \bibnamefont  [1]{#1}%
\providecommand \bibfnamefont [1]{#1}%
\providecommand \citenamefont [1]{#1}%
\providecommand \href@noop [0]{\@secondoftwo}%
\providecommand \href [0]{\begingroup \@sanitize@url \@href}%
\providecommand \@href[1]{\@@startlink{#1}\@@href}%
\providecommand \@@href[1]{\endgroup#1\@@endlink}%
\providecommand \@sanitize@url [0]{\catcode `\\12\catcode `\$12\catcode
  `\&12\catcode `\#12\catcode `\^12\catcode `\_12\catcode `\%12\relax}%
\providecommand \@@startlink[1]{}%
\providecommand \@@endlink[0]{}%
\providecommand \url  [0]{\begingroup\@sanitize@url \@url }%
\providecommand \@url [1]{\endgroup\@href {#1}{\urlprefix }}%
\providecommand \urlprefix  [0]{URL }%
\providecommand \Eprint [0]{\href }%
\providecommand \doibase [0]{http://dx.doi.org/}%
\providecommand \selectlanguage [0]{\@gobble}%
\providecommand \bibinfo  [0]{\@secondoftwo}%
\providecommand \bibfield  [0]{\@secondoftwo}%
\providecommand \translation [1]{[#1]}%
\providecommand \BibitemOpen [0]{}%
\providecommand \bibitemStop [0]{}%
\providecommand \bibitemNoStop [0]{.\EOS\space}%
\providecommand \EOS [0]{\spacefactor3000\relax}%
\providecommand \BibitemShut  [1]{\csname bibitem#1\endcsname}%
\let\auto@bib@innerbib\@empty
%</preamble>
\bibitem [{\citenamefont {Brataas}\ \emph {et~al.}(2012)\citenamefont
  {Brataas}, \citenamefont {Kent},\ and\ \citenamefont {Ohno}}]{Brataas2012}%
  \BibitemOpen
  \bibfield  {author} {\bibinfo {author} {\bibfnamefont {A.}~\bibnamefont
  {Brataas}}, \bibinfo {author} {\bibfnamefont {A.~D.}\ \bibnamefont {Kent}}, \
  and\ \bibinfo {author} {\bibfnamefont {H.}~\bibnamefont {Ohno}},\ }\href
  {\doibase 10.1038/nmat3311} {\bibfield  {journal} {\bibinfo  {journal} {Nat.
  Mater.}\ }\textbf {\bibinfo {volume} {11}},\ \bibinfo {pages} {372} (\bibinfo
  {year} {2012})}\BibitemShut {NoStop}%
\bibitem [{\citenamefont {Kent}\ and\ \citenamefont
  {Worledge}(2015)}]{Kent2015}%
  \BibitemOpen
  \bibfield  {author} {\bibinfo {author} {\bibfnamefont {A.~D.}\ \bibnamefont
  {Kent}}\ and\ \bibinfo {author} {\bibfnamefont {D.~C.}\ \bibnamefont
  {Worledge}},\ }\href {\doibase 10.1038/nnano.2015.24} {\bibfield  {journal}
  {\bibinfo  {journal} {Nat. Nanotechnol.}\ }\textbf {\bibinfo {volume} {10}},\
  \bibinfo {pages} {187} (\bibinfo {year} {2015})}\BibitemShut {NoStop}%
\bibitem [{\citenamefont {Bernevig}\ and\ \citenamefont
  {Zhang}(2005)}]{Bernevig2005}%
  \BibitemOpen
  \bibfield  {author} {\bibinfo {author} {\bibfnamefont {B.~A.}\ \bibnamefont
  {Bernevig}}\ and\ \bibinfo {author} {\bibfnamefont {S.-C.}\ \bibnamefont
  {Zhang}},\ }\href {\doibase 10.1103/PhysRevLett.95.016801} {\bibfield
  {journal} {\bibinfo  {journal} {Phys. Rev. Lett.}\ }\textbf {\bibinfo
  {volume} {95}},\ \bibinfo {pages} {016801} (\bibinfo {year} {2005})},\
  \Eprint {http://arxiv.org/abs/0411457} {arXiv:0411457 [cond-mat]}
  \BibitemShut {NoStop}%
\bibitem [{\citenamefont {Manchon}\ and\ \citenamefont
  {Zhang}(2008)}]{Manchon2008}%
  \BibitemOpen
  \bibfield  {author} {\bibinfo {author} {\bibfnamefont {A.}~\bibnamefont
  {Manchon}}\ and\ \bibinfo {author} {\bibfnamefont {S.}~\bibnamefont
  {Zhang}},\ }\href {\doibase 10.1103/PhysRevB.78.212405} {\bibfield  {journal}
  {\bibinfo  {journal} {Phys. Rev. B}\ }\textbf {\bibinfo {volume} {78}},\
  \bibinfo {pages} {212405} (\bibinfo {year} {2008})}\BibitemShut {NoStop}%
\bibitem [{\citenamefont {Manchon}\ and\ \citenamefont
  {Zhang}(2009)}]{Manchon2009}%
  \BibitemOpen
  \bibfield  {author} {\bibinfo {author} {\bibfnamefont {A.}~\bibnamefont
  {Manchon}}\ and\ \bibinfo {author} {\bibfnamefont {S.}~\bibnamefont
  {Zhang}},\ }\href {\doibase 10.1103/PhysRevB.79.094422} {\bibfield  {journal}
  {\bibinfo  {journal} {Phys. Rev. B}\ }\textbf {\bibinfo {volume} {79}},\
  \bibinfo {pages} {094422} (\bibinfo {year} {2009})}\BibitemShut {NoStop}%
\bibitem [{\citenamefont {Garate}\ and\ \citenamefont
  {MacDonald}(2009)}]{Garate2009}%
  \BibitemOpen
  \bibfield  {author} {\bibinfo {author} {\bibfnamefont {I.}~\bibnamefont
  {Garate}}\ and\ \bibinfo {author} {\bibfnamefont {A.~H.}\ \bibnamefont
  {MacDonald}},\ }\href {\doibase 10.1103/PhysRevB.80.134403} {\bibfield
  {journal} {\bibinfo  {journal} {Phys. Rev. B}\ }\textbf {\bibinfo {volume}
  {80}},\ \bibinfo {pages} {134403} (\bibinfo {year} {2009})}\BibitemShut
  {NoStop}%
\bibitem [{\citenamefont {Brataas}\ and\ \citenamefont
  {Hals}(2014)}]{Brataas2014}%
  \BibitemOpen
  \bibfield  {author} {\bibinfo {author} {\bibfnamefont {A.}~\bibnamefont
  {Brataas}}\ and\ \bibinfo {author} {\bibfnamefont {K.~M.~D.}\ \bibnamefont
  {Hals}},\ }\href {\doibase 10.1038/nnano.2014.8} {\bibfield  {journal}
  {\bibinfo  {journal} {Nat. Nanotechnol.}\ }\textbf {\bibinfo {volume} {9}},\
  \bibinfo {pages} {86} (\bibinfo {year} {2014})}\BibitemShut {NoStop}%
\bibitem [{\citenamefont {Chernyshov}\ \emph {et~al.}(2009)\citenamefont
  {Chernyshov}, \citenamefont {Overby}, \citenamefont {Liu}, \citenamefont
  {Furdyna}, \citenamefont {Lyanda-Geller},\ and\ \citenamefont
  {Rokhinson}}]{Chernyshov2008}%
  \BibitemOpen
  \bibfield  {author} {\bibinfo {author} {\bibfnamefont {A.}~\bibnamefont
  {Chernyshov}}, \bibinfo {author} {\bibfnamefont {M.}~\bibnamefont {Overby}},
  \bibinfo {author} {\bibfnamefont {X.}~\bibnamefont {Liu}}, \bibinfo {author}
  {\bibfnamefont {J.~K.}\ \bibnamefont {Furdyna}}, \bibinfo {author}
  {\bibfnamefont {Y.}~\bibnamefont {Lyanda-Geller}}, \ and\ \bibinfo {author}
  {\bibfnamefont {L.~P.}\ \bibnamefont {Rokhinson}},\ }\href {\doibase
  10.1038/nphys1362} {\bibfield  {journal} {\bibinfo  {journal} {Nat. Phys.}\
  }\textbf {\bibinfo {volume} {5}},\ \bibinfo {pages} {656} (\bibinfo {year}
  {2009})},\ \Eprint {http://arxiv.org/abs/0812.3160} {arXiv:0812.3160}
  \BibitemShut {NoStop}%
\bibitem [{\citenamefont {{Mihai Miron}}\ \emph {et~al.}(2010)\citenamefont
  {{Mihai Miron}}, \citenamefont {Gaudin}, \citenamefont {Auffret},
  \citenamefont {Rodmacq}, \citenamefont {Schuhl}, \citenamefont {Pizzini},
  \citenamefont {Vogel},\ and\ \citenamefont {Gambardella}}]{MihaiMiron2010}%
  \BibitemOpen
  \bibfield  {author} {\bibinfo {author} {\bibfnamefont {I.}~\bibnamefont
  {{Mihai Miron}}}, \bibinfo {author} {\bibfnamefont {G.}~\bibnamefont
  {Gaudin}}, \bibinfo {author} {\bibfnamefont {S.}~\bibnamefont {Auffret}},
  \bibinfo {author} {\bibfnamefont {B.}~\bibnamefont {Rodmacq}}, \bibinfo
  {author} {\bibfnamefont {A.}~\bibnamefont {Schuhl}}, \bibinfo {author}
  {\bibfnamefont {S.}~\bibnamefont {Pizzini}}, \bibinfo {author} {\bibfnamefont
  {J.}~\bibnamefont {Vogel}}, \ and\ \bibinfo {author} {\bibfnamefont
  {P.}~\bibnamefont {Gambardella}},\ }\href {\doibase 10.1038/nmat2613}
  {\bibfield  {journal} {\bibinfo  {journal} {Nat. Mater.}\ }\textbf {\bibinfo
  {volume} {9}},\ \bibinfo {pages} {230} (\bibinfo {year} {2010})}\BibitemShut
  {NoStop}%
\bibitem [{\citenamefont {Liu}\ \emph {et~al.}(2012{\natexlab{a}})\citenamefont
  {Liu}, \citenamefont {Pai}, \citenamefont {Li}, \citenamefont {Tseng},
  \citenamefont {Ralph},\ and\ \citenamefont {Buhrman}}]{Liu2012}%
  \BibitemOpen
  \bibfield  {author} {\bibinfo {author} {\bibfnamefont {L.}~\bibnamefont
  {Liu}}, \bibinfo {author} {\bibfnamefont {C.-F.}\ \bibnamefont {Pai}},
  \bibinfo {author} {\bibfnamefont {Y.}~\bibnamefont {Li}}, \bibinfo {author}
  {\bibfnamefont {H.~W.}\ \bibnamefont {Tseng}}, \bibinfo {author}
  {\bibfnamefont {D.~C.}\ \bibnamefont {Ralph}}, \ and\ \bibinfo {author}
  {\bibfnamefont {R.~A.}\ \bibnamefont {Buhrman}},\ }\href {\doibase
  10.1126/science.1218197} {\bibfield  {journal} {\bibinfo  {journal} {Science
  (80).}\ }\textbf {\bibinfo {volume} {336}},\ \bibinfo {pages} {555} (\bibinfo
  {year} {2012}{\natexlab{a}})},\ \Eprint {http://arxiv.org/abs/1203.2875}
  {arXiv:1203.2875} \BibitemShut {NoStop}%
\bibitem [{\citenamefont {Garello}\ \emph {et~al.}(2014)\citenamefont
  {Garello}, \citenamefont {Avci}, \citenamefont {Miron}, \citenamefont
  {Baumgartner}, \citenamefont {Ghosh}, \citenamefont {Auffret}, \citenamefont
  {Boulle}, \citenamefont {Gaudin},\ and\ \citenamefont
  {Gambardella}}]{Garello2014}%
  \BibitemOpen
  \bibfield  {author} {\bibinfo {author} {\bibfnamefont {K.}~\bibnamefont
  {Garello}}, \bibinfo {author} {\bibfnamefont {C.~O.}\ \bibnamefont {Avci}},
  \bibinfo {author} {\bibfnamefont {I.~M.}\ \bibnamefont {Miron}}, \bibinfo
  {author} {\bibfnamefont {M.}~\bibnamefont {Baumgartner}}, \bibinfo {author}
  {\bibfnamefont {A.}~\bibnamefont {Ghosh}}, \bibinfo {author} {\bibfnamefont
  {S.}~\bibnamefont {Auffret}}, \bibinfo {author} {\bibfnamefont
  {O.}~\bibnamefont {Boulle}}, \bibinfo {author} {\bibfnamefont
  {G.}~\bibnamefont {Gaudin}}, \ and\ \bibinfo {author} {\bibfnamefont
  {P.}~\bibnamefont {Gambardella}},\ }\href {\doibase 10.1063/1.4902443}
  {\bibfield  {journal} {\bibinfo  {journal} {Appl. Phys. Lett.}\ }\textbf
  {\bibinfo {volume} {105}},\ \bibinfo {pages} {212402} (\bibinfo {year}
  {2014})},\ \Eprint {http://arxiv.org/abs/1310.5586} {arXiv:1310.5586}
  \BibitemShut {NoStop}%
\bibitem [{\citenamefont {Miron}\ \emph
  {et~al.}(2011{\natexlab{a}})\citenamefont {Miron}, \citenamefont {Moore},
  \citenamefont {Szambolics}, \citenamefont {Buda-Prejbeanu}, \citenamefont
  {Auffret}, \citenamefont {Rodmacq}, \citenamefont {Pizzini}, \citenamefont
  {Vogel}, \citenamefont {Bonfim}, \citenamefont {Schuhl},\ and\ \citenamefont
  {Gaudin}}]{Miron2011b}%
  \BibitemOpen
  \bibfield  {author} {\bibinfo {author} {\bibfnamefont {I.~M.}\ \bibnamefont
  {Miron}}, \bibinfo {author} {\bibfnamefont {T.}~\bibnamefont {Moore}},
  \bibinfo {author} {\bibfnamefont {H.}~\bibnamefont {Szambolics}}, \bibinfo
  {author} {\bibfnamefont {L.~D.}\ \bibnamefont {Buda-Prejbeanu}}, \bibinfo
  {author} {\bibfnamefont {S.}~\bibnamefont {Auffret}}, \bibinfo {author}
  {\bibfnamefont {B.}~\bibnamefont {Rodmacq}}, \bibinfo {author} {\bibfnamefont
  {S.}~\bibnamefont {Pizzini}}, \bibinfo {author} {\bibfnamefont
  {J.}~\bibnamefont {Vogel}}, \bibinfo {author} {\bibfnamefont
  {M.}~\bibnamefont {Bonfim}}, \bibinfo {author} {\bibfnamefont
  {A.}~\bibnamefont {Schuhl}}, \ and\ \bibinfo {author} {\bibfnamefont
  {G.}~\bibnamefont {Gaudin}},\ }\href {\doibase 10.1038/nmat3020} {\bibfield
  {journal} {\bibinfo  {journal} {Nat. Mater.}\ }\textbf {\bibinfo {volume}
  {10}},\ \bibinfo {pages} {419} (\bibinfo {year} {2011}{\natexlab{a}})},\
  \Eprint {http://arxiv.org/abs/1302.2257} {arXiv:1302.2257} \BibitemShut
  {NoStop}%
\bibitem [{\citenamefont {Emori}\ \emph {et~al.}(2013)\citenamefont {Emori},
  \citenamefont {Bauer}, \citenamefont {Ahn}, \citenamefont {Martinez},\ and\
  \citenamefont {Beach}}]{Emori2013}%
  \BibitemOpen
  \bibfield  {author} {\bibinfo {author} {\bibfnamefont {S.}~\bibnamefont
  {Emori}}, \bibinfo {author} {\bibfnamefont {U.}~\bibnamefont {Bauer}},
  \bibinfo {author} {\bibfnamefont {S.-M.}\ \bibnamefont {Ahn}}, \bibinfo
  {author} {\bibfnamefont {E.}~\bibnamefont {Martinez}}, \ and\ \bibinfo
  {author} {\bibfnamefont {G.~S.~D.}\ \bibnamefont {Beach}},\ }\href {\doibase
  10.1038/nmat3675} {\bibfield  {journal} {\bibinfo  {journal} {Nat. Mater.}\
  }\textbf {\bibinfo {volume} {12}},\ \bibinfo {pages} {611} (\bibinfo {year}
  {2013})},\ \Eprint {http://arxiv.org/abs/1302.2257} {arXiv:1302.2257}
  \BibitemShut {NoStop}%
\bibitem [{\citenamefont {Yang}\ \emph {et~al.}(2015)\citenamefont {Yang},
  \citenamefont {Ryu},\ and\ \citenamefont {Parkin}}]{Yang2015}%
  \BibitemOpen
  \bibfield  {author} {\bibinfo {author} {\bibfnamefont {S.-H.}\ \bibnamefont
  {Yang}}, \bibinfo {author} {\bibfnamefont {K.-S.}\ \bibnamefont {Ryu}}, \
  and\ \bibinfo {author} {\bibfnamefont {S.}~\bibnamefont {Parkin}},\ }\href
  {\doibase 10.1038/nnano.2014.324} {\bibfield  {journal} {\bibinfo  {journal}
  {Nat. Nanotechnol.}\ }\textbf {\bibinfo {volume} {10}},\ \bibinfo {pages}
  {221} (\bibinfo {year} {2015})}\BibitemShut {NoStop}%
\bibitem [{\citenamefont {Liu}\ \emph {et~al.}(2012{\natexlab{b}})\citenamefont
  {Liu}, \citenamefont {Lee}, \citenamefont {Gudmundsen}, \citenamefont
  {Ralph},\ and\ \citenamefont {Buhrman}}]{Liu2012a}%
  \BibitemOpen
  \bibfield  {author} {\bibinfo {author} {\bibfnamefont {L.}~\bibnamefont
  {Liu}}, \bibinfo {author} {\bibfnamefont {O.~J.}\ \bibnamefont {Lee}},
  \bibinfo {author} {\bibfnamefont {T.~J.}\ \bibnamefont {Gudmundsen}},
  \bibinfo {author} {\bibfnamefont {D.~C.}\ \bibnamefont {Ralph}}, \ and\
  \bibinfo {author} {\bibfnamefont {R.~A.}\ \bibnamefont {Buhrman}},\ }\href
  {\doibase 10.1103/PhysRevLett.109.096602} {\bibfield  {journal} {\bibinfo
  {journal} {Phys. Rev. Lett.}\ }\textbf {\bibinfo {volume} {109}},\ \bibinfo
  {pages} {096602} (\bibinfo {year} {2012}{\natexlab{b}})},\ \Eprint
  {http://arxiv.org/abs/1110.6846} {arXiv:1110.6846} \BibitemShut {NoStop}%
\bibitem [{\citenamefont {Demidov}\ \emph {et~al.}(2012)\citenamefont
  {Demidov}, \citenamefont {Urazhdin}, \citenamefont {Ulrichs}, \citenamefont
  {Tiberkevich}, \citenamefont {Slavin}, \citenamefont {Baither}, \citenamefont
  {Schmitz},\ and\ \citenamefont {Demokritov}}]{Demidov2012}%
  \BibitemOpen
  \bibfield  {author} {\bibinfo {author} {\bibfnamefont {V.~E.}\ \bibnamefont
  {Demidov}}, \bibinfo {author} {\bibfnamefont {S.}~\bibnamefont {Urazhdin}},
  \bibinfo {author} {\bibfnamefont {H.}~\bibnamefont {Ulrichs}}, \bibinfo
  {author} {\bibfnamefont {V.}~\bibnamefont {Tiberkevich}}, \bibinfo {author}
  {\bibfnamefont {A.}~\bibnamefont {Slavin}}, \bibinfo {author} {\bibfnamefont
  {D.}~\bibnamefont {Baither}}, \bibinfo {author} {\bibfnamefont
  {G.}~\bibnamefont {Schmitz}}, \ and\ \bibinfo {author} {\bibfnamefont
  {S.~O.}\ \bibnamefont {Demokritov}},\ }\href {\doibase 10.1038/nmat3459}
  {\bibfield  {journal} {\bibinfo  {journal} {Nat. Mater.}\ }\textbf {\bibinfo
  {volume} {11}},\ \bibinfo {pages} {1028} (\bibinfo {year}
  {2012})}\BibitemShut {NoStop}%
\bibitem [{\citenamefont {Fukami}\ \emph
  {et~al.}(2016{\natexlab{a}})\citenamefont {Fukami}, \citenamefont {Zhang},
  \citenamefont {DuttaGupta}, \citenamefont {Kurenkov},\ and\ \citenamefont
  {Ohno}}]{Fukami2016}%
  \BibitemOpen
  \bibfield  {author} {\bibinfo {author} {\bibfnamefont {S.}~\bibnamefont
  {Fukami}}, \bibinfo {author} {\bibfnamefont {C.}~\bibnamefont {Zhang}},
  \bibinfo {author} {\bibfnamefont {S.}~\bibnamefont {DuttaGupta}}, \bibinfo
  {author} {\bibfnamefont {A.}~\bibnamefont {Kurenkov}}, \ and\ \bibinfo
  {author} {\bibfnamefont {H.}~\bibnamefont {Ohno}},\ }\href {\doibase
  10.1038/nmat4566} {\bibfield  {journal} {\bibinfo  {journal} {Nat. Mater.}\
  }\textbf {\bibinfo {volume} {15}},\ \bibinfo {pages} {535} (\bibinfo {year}
  {2016}{\natexlab{a}})},\ \Eprint {http://arxiv.org/abs/1507.00888}
  {arXiv:1507.00888} \BibitemShut {NoStop}%
\bibitem [{\citenamefont {Fukami}\ \emph
  {et~al.}(2016{\natexlab{b}})\citenamefont {Fukami}, \citenamefont {Anekawa},
  \citenamefont {Zhang},\ and\ \citenamefont {Ohno}}]{Fukami2016a}%
  \BibitemOpen
  \bibfield  {author} {\bibinfo {author} {\bibfnamefont {S.}~\bibnamefont
  {Fukami}}, \bibinfo {author} {\bibfnamefont {T.}~\bibnamefont {Anekawa}},
  \bibinfo {author} {\bibfnamefont {C.}~\bibnamefont {Zhang}}, \ and\ \bibinfo
  {author} {\bibfnamefont {H.}~\bibnamefont {Ohno}},\ }\href {\doibase
  10.1038/nnano.2016.29} {\bibfield  {journal} {\bibinfo  {journal} {Nat.
  Nanotechnol.}\ }\textbf {\bibinfo {volume} {11}},\ \bibinfo {pages} {621}
  (\bibinfo {year} {2016}{\natexlab{b}})}\BibitemShut {NoStop}%
\bibitem [{\citenamefont {Lau}\ \emph {et~al.}(2016)\citenamefont {Lau},
  \citenamefont {Betto}, \citenamefont {Rode}, \citenamefont {Coey},\ and\
  \citenamefont {Stamenov}}]{Lau2016}%
  \BibitemOpen
  \bibfield  {author} {\bibinfo {author} {\bibfnamefont {Y.-C.}\ \bibnamefont
  {Lau}}, \bibinfo {author} {\bibfnamefont {D.}~\bibnamefont {Betto}}, \bibinfo
  {author} {\bibfnamefont {K.}~\bibnamefont {Rode}}, \bibinfo {author}
  {\bibfnamefont {J.~M.~D.}\ \bibnamefont {Coey}}, \ and\ \bibinfo {author}
  {\bibfnamefont {P.}~\bibnamefont {Stamenov}},\ }\href {\doibase
  10.1038/nnano.2016.84} {\bibfield  {journal} {\bibinfo  {journal} {Nat.
  Nanotechnol.}\ }\textbf {\bibinfo {volume} {11}},\ \bibinfo {pages} {758}
  (\bibinfo {year} {2016})}\BibitemShut {NoStop}%
\bibitem [{\citenamefont {van~den Brink}\ \emph {et~al.}(2016)\citenamefont
  {van~den Brink}, \citenamefont {Vermijs}, \citenamefont {Solignac},
  \citenamefont {Koo}, \citenamefont {Kohlhepp}, \citenamefont {Swagten},\ and\
  \citenamefont {Koopmans}}]{VandenBrink2016}%
  \BibitemOpen
  \bibfield  {author} {\bibinfo {author} {\bibfnamefont {A.}~\bibnamefont
  {van~den Brink}}, \bibinfo {author} {\bibfnamefont {G.}~\bibnamefont
  {Vermijs}}, \bibinfo {author} {\bibfnamefont {A.}~\bibnamefont {Solignac}},
  \bibinfo {author} {\bibfnamefont {J.}~\bibnamefont {Koo}}, \bibinfo {author}
  {\bibfnamefont {J.~T.}\ \bibnamefont {Kohlhepp}}, \bibinfo {author}
  {\bibfnamefont {H.~J.~M.}\ \bibnamefont {Swagten}}, \ and\ \bibinfo {author}
  {\bibfnamefont {B.}~\bibnamefont {Koopmans}},\ }\href {\doibase
  10.1038/ncomms10854} {\bibfield  {journal} {\bibinfo  {journal} {Nat.
  Commun.}\ }\textbf {\bibinfo {volume} {7}},\ \bibinfo {pages} {10854}
  (\bibinfo {year} {2016})},\ \Eprint {http://arxiv.org/abs/1509.08752}
  {arXiv:1509.08752} \BibitemShut {NoStop}%
\bibitem [{\citenamefont {Qiu}\ \emph {et~al.}(2016)\citenamefont {Qiu},
  \citenamefont {Legrand}, \citenamefont {He}, \citenamefont {Wu},
  \citenamefont {Yu}, \citenamefont {Ramaswamy}, \citenamefont {Manchon},\ and\
  \citenamefont {Yang}}]{Qiu2016}%
  \BibitemOpen
  \bibfield  {author} {\bibinfo {author} {\bibfnamefont {X.}~\bibnamefont
  {Qiu}}, \bibinfo {author} {\bibfnamefont {W.}~\bibnamefont {Legrand}},
  \bibinfo {author} {\bibfnamefont {P.}~\bibnamefont {He}}, \bibinfo {author}
  {\bibfnamefont {Y.}~\bibnamefont {Wu}}, \bibinfo {author} {\bibfnamefont
  {J.}~\bibnamefont {Yu}}, \bibinfo {author} {\bibfnamefont {R.}~\bibnamefont
  {Ramaswamy}}, \bibinfo {author} {\bibfnamefont {A.}~\bibnamefont {Manchon}},
  \ and\ \bibinfo {author} {\bibfnamefont {H.}~\bibnamefont {Yang}},\ }\href
  {\doibase 10.1103/PhysRevLett.117.217206} {\bibfield  {journal} {\bibinfo
  {journal} {Phys. Rev. Lett.}\ }\textbf {\bibinfo {volume} {117}},\ \bibinfo
  {pages} {217206} (\bibinfo {year} {2016})},\ \Eprint
  {http://arxiv.org/abs/1610.06989} {arXiv:1610.06989} \BibitemShut {NoStop}%
\bibitem [{\citenamefont {Garello}\ \emph {et~al.}(2013)\citenamefont
  {Garello}, \citenamefont {Miron}, \citenamefont {Avci}, \citenamefont
  {Freimuth}, \citenamefont {Mokrousov}, \citenamefont {Bl{\"{u}}gel},
  \citenamefont {Auffret}, \citenamefont {Boulle}, \citenamefont {Gaudin},\
  and\ \citenamefont {Gambardella}}]{Garello2013}%
  \BibitemOpen
  \bibfield  {author} {\bibinfo {author} {\bibfnamefont {K.}~\bibnamefont
  {Garello}}, \bibinfo {author} {\bibfnamefont {I.~M.}\ \bibnamefont {Miron}},
  \bibinfo {author} {\bibfnamefont {C.~O.}\ \bibnamefont {Avci}}, \bibinfo
  {author} {\bibfnamefont {F.}~\bibnamefont {Freimuth}}, \bibinfo {author}
  {\bibfnamefont {Y.}~\bibnamefont {Mokrousov}}, \bibinfo {author}
  {\bibfnamefont {S.}~\bibnamefont {Bl{\"{u}}gel}}, \bibinfo {author}
  {\bibfnamefont {S.}~\bibnamefont {Auffret}}, \bibinfo {author} {\bibfnamefont
  {O.}~\bibnamefont {Boulle}}, \bibinfo {author} {\bibfnamefont
  {G.}~\bibnamefont {Gaudin}}, \ and\ \bibinfo {author} {\bibfnamefont
  {P.}~\bibnamefont {Gambardella}},\ }\href {\doibase 10.1038/nnano.2013.145}
  {\bibfield  {journal} {\bibinfo  {journal} {Nat. Nanotechnol.}\ }\textbf
  {\bibinfo {volume} {8}},\ \bibinfo {pages} {587} (\bibinfo {year} {2013})},\
  \Eprint {http://arxiv.org/abs/1301.3573} {arXiv:1301.3573} \BibitemShut
  {NoStop}%
\bibitem [{\citenamefont {Qiu}\ \emph {et~al.}(2015)\citenamefont {Qiu},
  \citenamefont {Deorani}, \citenamefont {Narayanapillai}, \citenamefont {Lee},
  \citenamefont {Lee}, \citenamefont {Lee},\ and\ \citenamefont
  {Yang}}]{Qiu2015b}%
  \BibitemOpen
  \bibfield  {author} {\bibinfo {author} {\bibfnamefont {X.}~\bibnamefont
  {Qiu}}, \bibinfo {author} {\bibfnamefont {P.}~\bibnamefont {Deorani}},
  \bibinfo {author} {\bibfnamefont {K.}~\bibnamefont {Narayanapillai}},
  \bibinfo {author} {\bibfnamefont {K.-S.}\ \bibnamefont {Lee}}, \bibinfo
  {author} {\bibfnamefont {K.-J.}\ \bibnamefont {Lee}}, \bibinfo {author}
  {\bibfnamefont {H.-W.}\ \bibnamefont {Lee}}, \ and\ \bibinfo {author}
  {\bibfnamefont {H.}~\bibnamefont {Yang}},\ }\href {\doibase
  10.1038/srep04491} {\bibfield  {journal} {\bibinfo  {journal} {Sci. Rep.}\
  }\textbf {\bibinfo {volume} {4}},\ \bibinfo {pages} {4491} (\bibinfo {year}
  {2015})},\ \Eprint {http://arxiv.org/abs/1404.1130} {arXiv:1404.1130}
  \BibitemShut {NoStop}%
\bibitem [{\citenamefont {Ando}\ \emph {et~al.}(2008)\citenamefont {Ando},
  \citenamefont {Takahashi}, \citenamefont {Harii}, \citenamefont {Sasage},
  \citenamefont {Ieda}, \citenamefont {Maekawa},\ and\ \citenamefont
  {Saitoh}}]{Ando2008}%
  \BibitemOpen
  \bibfield  {author} {\bibinfo {author} {\bibfnamefont {K.}~\bibnamefont
  {Ando}}, \bibinfo {author} {\bibfnamefont {S.}~\bibnamefont {Takahashi}},
  \bibinfo {author} {\bibfnamefont {K.}~\bibnamefont {Harii}}, \bibinfo
  {author} {\bibfnamefont {K.}~\bibnamefont {Sasage}}, \bibinfo {author}
  {\bibfnamefont {J.}~\bibnamefont {Ieda}}, \bibinfo {author} {\bibfnamefont
  {S.}~\bibnamefont {Maekawa}}, \ and\ \bibinfo {author} {\bibfnamefont
  {E.}~\bibnamefont {Saitoh}},\ }\href {\doibase
  10.1103/PhysRevLett.101.036601} {\bibfield  {journal} {\bibinfo  {journal}
  {Phys. Rev. Lett.}\ }\textbf {\bibinfo {volume} {101}},\ \bibinfo {pages}
  {036601} (\bibinfo {year} {2008})}\BibitemShut {NoStop}%
\bibitem [{\citenamefont {Haney}\ \emph
  {et~al.}(2013{\natexlab{a}})\citenamefont {Haney}, \citenamefont {Lee},
  \citenamefont {Lee}, \citenamefont {Manchon},\ and\ \citenamefont
  {Stiles}}]{Haney2013}%
  \BibitemOpen
  \bibfield  {author} {\bibinfo {author} {\bibfnamefont {P.~M.}\ \bibnamefont
  {Haney}}, \bibinfo {author} {\bibfnamefont {H.-W.}\ \bibnamefont {Lee}},
  \bibinfo {author} {\bibfnamefont {K.-J.}\ \bibnamefont {Lee}}, \bibinfo
  {author} {\bibfnamefont {A.}~\bibnamefont {Manchon}}, \ and\ \bibinfo
  {author} {\bibfnamefont {M.~D.}\ \bibnamefont {Stiles}},\ }\href {\doibase
  10.1103/PhysRevB.87.174411} {\bibfield  {journal} {\bibinfo  {journal} {Phys.
  Rev. B}\ }\textbf {\bibinfo {volume} {87}},\ \bibinfo {pages} {174411}
  (\bibinfo {year} {2013}{\natexlab{a}})},\ \Eprint
  {http://arxiv.org/abs/1301.4513} {arXiv:1301.4513} \BibitemShut {NoStop}%
\bibitem [{\citenamefont {Liu}\ \emph {et~al.}(2011)\citenamefont {Liu},
  \citenamefont {Moriyama}, \citenamefont {Ralph},\ and\ \citenamefont
  {Buhrman}}]{Liu2011c}%
  \BibitemOpen
  \bibfield  {author} {\bibinfo {author} {\bibfnamefont {L.}~\bibnamefont
  {Liu}}, \bibinfo {author} {\bibfnamefont {T.}~\bibnamefont {Moriyama}},
  \bibinfo {author} {\bibfnamefont {D.~C.}\ \bibnamefont {Ralph}}, \ and\
  \bibinfo {author} {\bibfnamefont {R.~A.}\ \bibnamefont {Buhrman}},\ }\href
  {\doibase 10.1103/PhysRevLett.106.036601} {\bibfield  {journal} {\bibinfo
  {journal} {Phys. Rev. Lett.}\ }\textbf {\bibinfo {volume} {106}},\ \bibinfo
  {pages} {036601} (\bibinfo {year} {2011})},\ \Eprint
  {http://arxiv.org/abs/1011.2788} {arXiv:1011.2788} \BibitemShut {NoStop}%
\bibitem [{\citenamefont {Ivchenko}\ and\ \citenamefont
  {Pikus}(1978)}]{Ivchenko1978}%
  \BibitemOpen
  \bibfield  {author} {\bibinfo {author} {\bibfnamefont {E.~L.}\ \bibnamefont
  {Ivchenko}}\ and\ \bibinfo {author} {\bibfnamefont {G.~E.}\ \bibnamefont
  {Pikus}},\ }\href {http://www.jetpletters.ac.ru/ps/1554/article_23792.shtml}
  {\bibfield  {journal} {\bibinfo  {journal} {JETP Lett.}\ }\textbf {\bibinfo
  {volume} {27}},\ \bibinfo {pages} {604} (\bibinfo {year} {1978})}\BibitemShut
  {NoStop}%
\bibitem [{\citenamefont {Edelstein}(1990)}]{Edelstein1990}%
  \BibitemOpen
  \bibfield  {author} {\bibinfo {author} {\bibfnamefont {V.}~\bibnamefont
  {Edelstein}},\ }\href {\doibase 10.1016/0038-1098(90)90963-C} {\bibfield
  {journal} {\bibinfo  {journal} {Solid State Commun.}\ }\textbf {\bibinfo
  {volume} {73}},\ \bibinfo {pages} {233} (\bibinfo {year} {1990})}\BibitemShut
  {NoStop}%
\bibitem [{\citenamefont {Haney}\ \emph
  {et~al.}(2013{\natexlab{b}})\citenamefont {Haney}, \citenamefont {Lee},
  \citenamefont {Lee}, \citenamefont {Manchon},\ and\ \citenamefont
  {Stiles}}]{Haney2013a}%
  \BibitemOpen
  \bibfield  {author} {\bibinfo {author} {\bibfnamefont {P.~M.}\ \bibnamefont
  {Haney}}, \bibinfo {author} {\bibfnamefont {H.-W.}\ \bibnamefont {Lee}},
  \bibinfo {author} {\bibfnamefont {K.-J.}\ \bibnamefont {Lee}}, \bibinfo
  {author} {\bibfnamefont {A.}~\bibnamefont {Manchon}}, \ and\ \bibinfo
  {author} {\bibfnamefont {M.~D.}\ \bibnamefont {Stiles}},\ }\href {\doibase
  10.1103/PhysRevB.88.214417} {\bibfield  {journal} {\bibinfo  {journal} {Phys.
  Rev. B}\ }\textbf {\bibinfo {volume} {88}},\ \bibinfo {pages} {214417}
  (\bibinfo {year} {2013}{\natexlab{b}})},\ \Eprint
  {http://arxiv.org/abs/1301.4513v1} {arXiv:1301.4513v1} \BibitemShut {NoStop}%
\bibitem [{\citenamefont {Fan}\ \emph {et~al.}(2013)\citenamefont {Fan},
  \citenamefont {Wu}, \citenamefont {Chen}, \citenamefont {Jerry},
  \citenamefont {Zhang},\ and\ \citenamefont {Xiao}}]{Fan2013}%
  \BibitemOpen
  \bibfield  {author} {\bibinfo {author} {\bibfnamefont {X.}~\bibnamefont
  {Fan}}, \bibinfo {author} {\bibfnamefont {J.}~\bibnamefont {Wu}}, \bibinfo
  {author} {\bibfnamefont {Y.}~\bibnamefont {Chen}}, \bibinfo {author}
  {\bibfnamefont {M.~J.}\ \bibnamefont {Jerry}}, \bibinfo {author}
  {\bibfnamefont {H.}~\bibnamefont {Zhang}}, \ and\ \bibinfo {author}
  {\bibfnamefont {J.~Q.}\ \bibnamefont {Xiao}},\ }\href {\doibase
  10.1038/ncomms2709} {\bibfield  {journal} {\bibinfo  {journal} {Nat.
  Commun.}\ }\textbf {\bibinfo {volume} {4}},\ \bibinfo {pages} {1799}
  (\bibinfo {year} {2013})}\BibitemShut {NoStop}%
\bibitem [{\citenamefont {Pai}\ \emph {et~al.}(2014)\citenamefont {Pai},
  \citenamefont {Nguyen}, \citenamefont {Belvin}, \citenamefont
  {Vilela-Le{\~{a}}o}, \citenamefont {Ralph},\ and\ \citenamefont
  {Buhrman}}]{Pai2014}%
  \BibitemOpen
  \bibfield  {author} {\bibinfo {author} {\bibfnamefont {C.-F.}\ \bibnamefont
  {Pai}}, \bibinfo {author} {\bibfnamefont {M.-H.}\ \bibnamefont {Nguyen}},
  \bibinfo {author} {\bibfnamefont {C.}~\bibnamefont {Belvin}}, \bibinfo
  {author} {\bibfnamefont {L.~H.}\ \bibnamefont {Vilela-Le{\~{a}}o}}, \bibinfo
  {author} {\bibfnamefont {D.~C.}\ \bibnamefont {Ralph}}, \ and\ \bibinfo
  {author} {\bibfnamefont {R.~A.}\ \bibnamefont {Buhrman}},\ }\href {\doibase
  10.1063/1.4866965} {\bibfield  {journal} {\bibinfo  {journal} {Appl. Phys.
  Lett.}\ }\textbf {\bibinfo {volume} {104}},\ \bibinfo {pages} {082407}
  (\bibinfo {year} {2014})},\ \Eprint {http://arxiv.org/abs/1401.4617}
  {arXiv:1401.4617} \BibitemShut {NoStop}%
\bibitem [{\citenamefont {Kim}\ \emph {et~al.}(2012)\citenamefont {Kim},
  \citenamefont {Sinha}, \citenamefont {Hayashi}, \citenamefont {Yamanouchi},
  \citenamefont {Fukami}, \citenamefont {Suzuki}, \citenamefont {Mitani},\ and\
  \citenamefont {Ohno}}]{Kim2013a}%
  \BibitemOpen
  \bibfield  {author} {\bibinfo {author} {\bibfnamefont {J.}~\bibnamefont
  {Kim}}, \bibinfo {author} {\bibfnamefont {J.}~\bibnamefont {Sinha}}, \bibinfo
  {author} {\bibfnamefont {M.}~\bibnamefont {Hayashi}}, \bibinfo {author}
  {\bibfnamefont {M.}~\bibnamefont {Yamanouchi}}, \bibinfo {author}
  {\bibfnamefont {S.}~\bibnamefont {Fukami}}, \bibinfo {author} {\bibfnamefont
  {T.}~\bibnamefont {Suzuki}}, \bibinfo {author} {\bibfnamefont
  {S.}~\bibnamefont {Mitani}}, \ and\ \bibinfo {author} {\bibfnamefont
  {H.}~\bibnamefont {Ohno}},\ }\href {\doibase 10.1038/nmat3522} {\bibfield
  {journal} {\bibinfo  {journal} {Nat. Mater.}\ }\textbf {\bibinfo {volume}
  {12}},\ \bibinfo {pages} {240} (\bibinfo {year} {2012})},\ \Eprint
  {http://arxiv.org/abs/1207.2521} {arXiv:1207.2521} \BibitemShut {NoStop}%
\bibitem [{\citenamefont {Kim}\ \emph {et~al.}(2014)\citenamefont {Kim},
  \citenamefont {Sinha}, \citenamefont {Mitani}, \citenamefont {Hayashi},
  \citenamefont {Takahashi}, \citenamefont {Maekawa}, \citenamefont
  {Yamanouchi},\ and\ \citenamefont {Ohno}}]{Kim2014a}%
  \BibitemOpen
  \bibfield  {author} {\bibinfo {author} {\bibfnamefont {J.}~\bibnamefont
  {Kim}}, \bibinfo {author} {\bibfnamefont {J.}~\bibnamefont {Sinha}}, \bibinfo
  {author} {\bibfnamefont {S.}~\bibnamefont {Mitani}}, \bibinfo {author}
  {\bibfnamefont {M.}~\bibnamefont {Hayashi}}, \bibinfo {author} {\bibfnamefont
  {S.}~\bibnamefont {Takahashi}}, \bibinfo {author} {\bibfnamefont
  {S.}~\bibnamefont {Maekawa}}, \bibinfo {author} {\bibfnamefont
  {M.}~\bibnamefont {Yamanouchi}}, \ and\ \bibinfo {author} {\bibfnamefont
  {H.}~\bibnamefont {Ohno}},\ }\href {\doibase 10.1103/PhysRevB.89.174424}
  {\bibfield  {journal} {\bibinfo  {journal} {Phys. Rev. B}\ }\textbf {\bibinfo
  {volume} {89}},\ \bibinfo {pages} {174424} (\bibinfo {year} {2014})},\
  \Eprint {http://arxiv.org/abs/1402.6388} {arXiv:1402.6388} \BibitemShut
  {NoStop}%
\bibitem [{\citenamefont {Nguyen}\ \emph {et~al.}(2016)\citenamefont {Nguyen},
  \citenamefont {Ralph},\ and\ \citenamefont {Buhrman}}]{Nguyen2016}%
  \BibitemOpen
  \bibfield  {author} {\bibinfo {author} {\bibfnamefont {M.-H.}\ \bibnamefont
  {Nguyen}}, \bibinfo {author} {\bibfnamefont {D.~C.}\ \bibnamefont {Ralph}}, \
  and\ \bibinfo {author} {\bibfnamefont {R.~A.}\ \bibnamefont {Buhrman}},\
  }\href {\doibase 10.1103/PhysRevLett.116.126601} {\bibfield  {journal}
  {\bibinfo  {journal} {Phys. Rev. Lett.}\ }\textbf {\bibinfo {volume} {116}},\
  \bibinfo {pages} {126601} (\bibinfo {year} {2016})},\ \Eprint
  {http://arxiv.org/abs/1512.06931} {arXiv:1512.06931} \BibitemShut {NoStop}%
\bibitem [{\citenamefont {Freimuth}\ \emph {et~al.}(2014)\citenamefont
  {Freimuth}, \citenamefont {Bl{\"{u}}gel},\ and\ \citenamefont
  {Mokrousov}}]{Freimuth2014}%
  \BibitemOpen
  \bibfield  {author} {\bibinfo {author} {\bibfnamefont {F.}~\bibnamefont
  {Freimuth}}, \bibinfo {author} {\bibfnamefont {S.}~\bibnamefont
  {Bl{\"{u}}gel}}, \ and\ \bibinfo {author} {\bibfnamefont {Y.}~\bibnamefont
  {Mokrousov}},\ }\href {\doibase 10.1103/PhysRevB.90.174423} {\bibfield
  {journal} {\bibinfo  {journal} {Phys. Rev. B}\ }\textbf {\bibinfo {volume}
  {90}},\ \bibinfo {pages} {174423} (\bibinfo {year} {2014})},\ \Eprint
  {http://arxiv.org/abs/1305.4873} {arXiv:1305.4873} \BibitemShut {NoStop}%
\bibitem [{\citenamefont {Kurebayashi}\ \emph {et~al.}(2014)\citenamefont
  {Kurebayashi}, \citenamefont {Sinova}, \citenamefont {Fang}, \citenamefont
  {Irvine}, \citenamefont {Skinner}, \citenamefont {Wunderlich}, \citenamefont
  {Nov{\'{a}}k}, \citenamefont {Campion}, \citenamefont {Gallagher},
  \citenamefont {Vehstedt}, \citenamefont {Z{\^{a}}rbo}, \citenamefont
  {V{\'{y}}born{\'{y}}}, \citenamefont {Ferguson},\ and\ \citenamefont
  {Jungwirth}}]{Kurebayashi2014}%
  \BibitemOpen
  \bibfield  {author} {\bibinfo {author} {\bibfnamefont {H.}~\bibnamefont
  {Kurebayashi}}, \bibinfo {author} {\bibfnamefont {J.}~\bibnamefont {Sinova}},
  \bibinfo {author} {\bibfnamefont {D.}~\bibnamefont {Fang}}, \bibinfo {author}
  {\bibfnamefont {A.~C.}\ \bibnamefont {Irvine}}, \bibinfo {author}
  {\bibfnamefont {T.~D.}\ \bibnamefont {Skinner}}, \bibinfo {author}
  {\bibfnamefont {J.}~\bibnamefont {Wunderlich}}, \bibinfo {author}
  {\bibfnamefont {V.}~\bibnamefont {Nov{\'{a}}k}}, \bibinfo {author}
  {\bibfnamefont {R.~P.}\ \bibnamefont {Campion}}, \bibinfo {author}
  {\bibfnamefont {B.~L.}\ \bibnamefont {Gallagher}}, \bibinfo {author}
  {\bibfnamefont {E.~K.}\ \bibnamefont {Vehstedt}}, \bibinfo {author}
  {\bibfnamefont {L.~P.}\ \bibnamefont {Z{\^{a}}rbo}}, \bibinfo {author}
  {\bibfnamefont {K.}~\bibnamefont {V{\'{y}}born{\'{y}}}}, \bibinfo {author}
  {\bibfnamefont {A.~J.}\ \bibnamefont {Ferguson}}, \ and\ \bibinfo {author}
  {\bibfnamefont {T.}~\bibnamefont {Jungwirth}},\ }\href {\doibase
  10.1038/nnano.2014.15} {\bibfield  {journal} {\bibinfo  {journal} {Nat.
  Nanotechnol.}\ }\textbf {\bibinfo {volume} {9}},\ \bibinfo {pages} {211}
  (\bibinfo {year} {2014})}\BibitemShut {NoStop}%
\bibitem [{\citenamefont {Manchon}(2014)}]{Manchon2014}%
  \BibitemOpen
  \bibfield  {author} {\bibinfo {author} {\bibfnamefont {A.}~\bibnamefont
  {Manchon}},\ }\href {\doibase 10.1038/nphys2957} {\bibfield  {journal}
  {\bibinfo  {journal} {Nat. Phys.}\ }\textbf {\bibinfo {volume} {10}},\
  \bibinfo {pages} {340} (\bibinfo {year} {2014})}\BibitemShut {NoStop}%
\bibitem [{\citenamefont {Li}\ \emph {et~al.}(2015)\citenamefont {Li},
  \citenamefont {Gao}, \citenamefont {Z{\^{a}}rbo}, \citenamefont
  {V{\'{y}}born{\'{y}}}, \citenamefont {Wang}, \citenamefont {Garate},
  \citenamefont {Dogan}, \citenamefont {{\v{C}}ejchan}, \citenamefont {Sinova},
  \citenamefont {Jungwirth},\ and\ \citenamefont {Manchon}}]{Li2015}%
  \BibitemOpen
  \bibfield  {author} {\bibinfo {author} {\bibfnamefont {H.}~\bibnamefont
  {Li}}, \bibinfo {author} {\bibfnamefont {H.}~\bibnamefont {Gao}}, \bibinfo
  {author} {\bibfnamefont {L.~P.}\ \bibnamefont {Z{\^{a}}rbo}}, \bibinfo
  {author} {\bibfnamefont {K.}~\bibnamefont {V{\'{y}}born{\'{y}}}}, \bibinfo
  {author} {\bibfnamefont {X.}~\bibnamefont {Wang}}, \bibinfo {author}
  {\bibfnamefont {I.}~\bibnamefont {Garate}}, \bibinfo {author} {\bibfnamefont
  {F.}~\bibnamefont {Dogan}}, \bibinfo {author} {\bibfnamefont
  {A.}~\bibnamefont {{\v{C}}ejchan}}, \bibinfo {author} {\bibfnamefont
  {J.}~\bibnamefont {Sinova}}, \bibinfo {author} {\bibfnamefont
  {T.}~\bibnamefont {Jungwirth}}, \ and\ \bibinfo {author} {\bibfnamefont
  {A.}~\bibnamefont {Manchon}},\ }\href {\doibase 10.1103/PhysRevB.91.134402}
  {\bibfield  {journal} {\bibinfo  {journal} {Phys. Rev. B}\ }\textbf {\bibinfo
  {volume} {91}},\ \bibinfo {pages} {134402} (\bibinfo {year}
  {2015})}\BibitemShut {NoStop}%
\bibitem [{\citenamefont {Lifshits}\ and\ \citenamefont
  {Dyakonov}(2009)}]{Lifshits2009}%
  \BibitemOpen
  \bibfield  {author} {\bibinfo {author} {\bibfnamefont {M.~B.}\ \bibnamefont
  {Lifshits}}\ and\ \bibinfo {author} {\bibfnamefont {M.~I.}\ \bibnamefont
  {Dyakonov}},\ }\href {\doibase 10.1103/PhysRevLett.103.186601} {\bibfield
  {journal} {\bibinfo  {journal} {Phys. Rev. Lett.}\ }\textbf {\bibinfo
  {volume} {103}},\ \bibinfo {pages} {186601} (\bibinfo {year} {2009})},\
  \Eprint {http://arxiv.org/abs/0905.4469} {arXiv:0905.4469} \BibitemShut
  {NoStop}%
\bibitem [{\citenamefont {Saidaoui}\ and\ \citenamefont
  {Manchon}(2016)}]{Saidaoui2016}%
  \BibitemOpen
  \bibfield  {author} {\bibinfo {author} {\bibfnamefont {H.~B.~M.}\
  \bibnamefont {Saidaoui}}\ and\ \bibinfo {author} {\bibfnamefont
  {A.}~\bibnamefont {Manchon}},\ }\href {\doibase
  10.1103/PhysRevLett.117.036601} {\bibfield  {journal} {\bibinfo  {journal}
  {Phys. Rev. Lett.}\ }\textbf {\bibinfo {volume} {117}},\ \bibinfo {pages}
  {036601} (\bibinfo {year} {2016})},\ \Eprint
  {http://arxiv.org/abs/1511.03454} {arXiv:1511.03454} \BibitemShut {NoStop}%
\bibitem [{\citenamefont {Shiomi}\ \emph {et~al.}(2014)\citenamefont {Shiomi},
  \citenamefont {Nomura}, \citenamefont {Kajiwara}, \citenamefont {Eto},
  \citenamefont {Novak}, \citenamefont {Segawa}, \citenamefont {Ando},\ and\
  \citenamefont {Saitoh}}]{Shiomi2014}%
  \BibitemOpen
  \bibfield  {author} {\bibinfo {author} {\bibfnamefont {Y.}~\bibnamefont
  {Shiomi}}, \bibinfo {author} {\bibfnamefont {K.}~\bibnamefont {Nomura}},
  \bibinfo {author} {\bibfnamefont {Y.}~\bibnamefont {Kajiwara}}, \bibinfo
  {author} {\bibfnamefont {K.}~\bibnamefont {Eto}}, \bibinfo {author}
  {\bibfnamefont {M.}~\bibnamefont {Novak}}, \bibinfo {author} {\bibfnamefont
  {K.}~\bibnamefont {Segawa}}, \bibinfo {author} {\bibfnamefont
  {Y.}~\bibnamefont {Ando}}, \ and\ \bibinfo {author} {\bibfnamefont
  {E.}~\bibnamefont {Saitoh}},\ }\href {\doibase
  10.1103/PhysRevLett.113.196601} {\bibfield  {journal} {\bibinfo  {journal}
  {Phys. Rev. Lett.}\ }\textbf {\bibinfo {volume} {113}},\ \bibinfo {pages}
  {196601} (\bibinfo {year} {2014})}\BibitemShut {NoStop}%
\bibitem [{\citenamefont {Jamali}\ \emph {et~al.}(2015)\citenamefont {Jamali},
  \citenamefont {Lee}, \citenamefont {Jeong}, \citenamefont {Mahfouzi},
  \citenamefont {Lv}, \citenamefont {Zhao}, \citenamefont {Nikoli{\'{c}}},
  \citenamefont {Mkhoyan}, \citenamefont {Samarth},\ and\ \citenamefont
  {Wang}}]{Jamali2015}%
  \BibitemOpen
  \bibfield  {author} {\bibinfo {author} {\bibfnamefont {M.}~\bibnamefont
  {Jamali}}, \bibinfo {author} {\bibfnamefont {J.~S.}\ \bibnamefont {Lee}},
  \bibinfo {author} {\bibfnamefont {J.~S.}\ \bibnamefont {Jeong}}, \bibinfo
  {author} {\bibfnamefont {F.}~\bibnamefont {Mahfouzi}}, \bibinfo {author}
  {\bibfnamefont {Y.}~\bibnamefont {Lv}}, \bibinfo {author} {\bibfnamefont
  {Z.}~\bibnamefont {Zhao}}, \bibinfo {author} {\bibfnamefont {B.~K.}\
  \bibnamefont {Nikoli{\'{c}}}}, \bibinfo {author} {\bibfnamefont {K.~A.}\
  \bibnamefont {Mkhoyan}}, \bibinfo {author} {\bibfnamefont {N.}~\bibnamefont
  {Samarth}}, \ and\ \bibinfo {author} {\bibfnamefont {J.-P.}\ \bibnamefont
  {Wang}},\ }\href {\doibase 10.1021/acs.nanolett.5b03274} {\bibfield
  {journal} {\bibinfo  {journal} {Nano Lett.}\ }\textbf {\bibinfo {volume}
  {15}},\ \bibinfo {pages} {7126} (\bibinfo {year} {2015})}\BibitemShut
  {NoStop}%
\bibitem [{\citenamefont {Rojas-S{\'{a}}nchez}\ \emph
  {et~al.}(2016)\citenamefont {Rojas-S{\'{a}}nchez}, \citenamefont
  {Oyarz{\'{u}}n}, \citenamefont {Fu}, \citenamefont {Marty}, \citenamefont
  {Vergnaud}, \citenamefont {Gambarelli}, \citenamefont {Vila}, \citenamefont
  {Jamet}, \citenamefont {Ohtsubo}, \citenamefont {Taleb-Ibrahimi},
  \citenamefont {{Le F{\`{e}}vre}}, \citenamefont {Bertran}, \citenamefont
  {Reyren}, \citenamefont {George},\ and\ \citenamefont
  {Fert}}]{Rojas-Sanchez2016}%
  \BibitemOpen
  \bibfield  {author} {\bibinfo {author} {\bibfnamefont {J.-C.}\ \bibnamefont
  {Rojas-S{\'{a}}nchez}}, \bibinfo {author} {\bibfnamefont {S.}~\bibnamefont
  {Oyarz{\'{u}}n}}, \bibinfo {author} {\bibfnamefont {Y.}~\bibnamefont {Fu}},
  \bibinfo {author} {\bibfnamefont {A.}~\bibnamefont {Marty}}, \bibinfo
  {author} {\bibfnamefont {C.}~\bibnamefont {Vergnaud}}, \bibinfo {author}
  {\bibfnamefont {S.}~\bibnamefont {Gambarelli}}, \bibinfo {author}
  {\bibfnamefont {L.}~\bibnamefont {Vila}}, \bibinfo {author} {\bibfnamefont
  {M.}~\bibnamefont {Jamet}}, \bibinfo {author} {\bibfnamefont
  {Y.}~\bibnamefont {Ohtsubo}}, \bibinfo {author} {\bibfnamefont
  {A.}~\bibnamefont {Taleb-Ibrahimi}}, \bibinfo {author} {\bibfnamefont
  {P.}~\bibnamefont {{Le F{\`{e}}vre}}}, \bibinfo {author} {\bibfnamefont
  {F.}~\bibnamefont {Bertran}}, \bibinfo {author} {\bibfnamefont
  {N.}~\bibnamefont {Reyren}}, \bibinfo {author} {\bibfnamefont {J.-M.}\
  \bibnamefont {George}}, \ and\ \bibinfo {author} {\bibfnamefont
  {A.}~\bibnamefont {Fert}},\ }\href {\doibase 10.1103/PhysRevLett.116.096602}
  {\bibfield  {journal} {\bibinfo  {journal} {Phys. Rev. Lett.}\ }\textbf
  {\bibinfo {volume} {116}},\ \bibinfo {pages} {096602} (\bibinfo {year}
  {2016})}\BibitemShut {NoStop}%
\bibitem [{\citenamefont {Fan}\ \emph {et~al.}(2016{\natexlab{a}})\citenamefont
  {Fan}, \citenamefont {Kou}, \citenamefont {Upadhyaya}, \citenamefont {Shao},
  \citenamefont {Pan}, \citenamefont {Lang}, \citenamefont {Che}, \citenamefont
  {Tang}, \citenamefont {Montazeri}, \citenamefont {Murata}, \citenamefont
  {Chang}, \citenamefont {Akyol}, \citenamefont {Yu}, \citenamefont {Nie},
  \citenamefont {Wong}, \citenamefont {Liu}, \citenamefont {Wang},
  \citenamefont {Tserkovnyak},\ and\ \citenamefont {Wang}}]{Fan2016}%
  \BibitemOpen
  \bibfield  {author} {\bibinfo {author} {\bibfnamefont {Y.}~\bibnamefont
  {Fan}}, \bibinfo {author} {\bibfnamefont {X.}~\bibnamefont {Kou}}, \bibinfo
  {author} {\bibfnamefont {P.}~\bibnamefont {Upadhyaya}}, \bibinfo {author}
  {\bibfnamefont {Q.}~\bibnamefont {Shao}}, \bibinfo {author} {\bibfnamefont
  {L.}~\bibnamefont {Pan}}, \bibinfo {author} {\bibfnamefont {M.}~\bibnamefont
  {Lang}}, \bibinfo {author} {\bibfnamefont {X.}~\bibnamefont {Che}}, \bibinfo
  {author} {\bibfnamefont {J.}~\bibnamefont {Tang}}, \bibinfo {author}
  {\bibfnamefont {M.}~\bibnamefont {Montazeri}}, \bibinfo {author}
  {\bibfnamefont {K.}~\bibnamefont {Murata}}, \bibinfo {author} {\bibfnamefont
  {L.-T.}\ \bibnamefont {Chang}}, \bibinfo {author} {\bibfnamefont
  {M.}~\bibnamefont {Akyol}}, \bibinfo {author} {\bibfnamefont
  {G.}~\bibnamefont {Yu}}, \bibinfo {author} {\bibfnamefont {T.}~\bibnamefont
  {Nie}}, \bibinfo {author} {\bibfnamefont {K.~L.}\ \bibnamefont {Wong}},
  \bibinfo {author} {\bibfnamefont {J.}~\bibnamefont {Liu}}, \bibinfo {author}
  {\bibfnamefont {Y.}~\bibnamefont {Wang}}, \bibinfo {author} {\bibfnamefont
  {Y.}~\bibnamefont {Tserkovnyak}}, \ and\ \bibinfo {author} {\bibfnamefont
  {K.~L.}\ \bibnamefont {Wang}},\ }\href {\doibase 10.1038/nnano.2015.294}
  {\bibfield  {journal} {\bibinfo  {journal} {Nature Nanotechnology}\ }\textbf
  {\bibinfo {volume} {11}},\ \bibinfo {pages} {352} (\bibinfo {year}
  {2016}{\natexlab{a}})},\ \Eprint {http://arxiv.org/abs/1511.07442}
  {arXiv:1511.07442} \BibitemShut {NoStop}%
\bibitem [{\citenamefont {Qi}\ and\ \citenamefont {Zhang}(2011)}]{Qi2011}%
  \BibitemOpen
  \bibfield  {author} {\bibinfo {author} {\bibfnamefont {X.-L.}\ \bibnamefont
  {Qi}}\ and\ \bibinfo {author} {\bibfnamefont {S.-C.}\ \bibnamefont {Zhang}},\
  }\href {\doibase 10.1103/RevModPhys.83.1057} {\bibfield  {journal} {\bibinfo
  {journal} {Rev. Mod. Phys.}\ }\textbf {\bibinfo {volume} {83}},\ \bibinfo
  {pages} {1057} (\bibinfo {year} {2011})},\ \Eprint
  {http://arxiv.org/abs/1008.2026} {arXiv:1008.2026} \BibitemShut {NoStop}%
\bibitem [{\citenamefont {Mellnik}\ \emph {et~al.}(2014)\citenamefont
  {Mellnik}, \citenamefont {Lee}, \citenamefont {Richardella}, \citenamefont
  {Grab}, \citenamefont {Mintun}, \citenamefont {Fischer}, \citenamefont
  {Vaezi}, \citenamefont {Manchon}, \citenamefont {Kim}, \citenamefont
  {Samarth},\ and\ \citenamefont {Ralph}}]{Mellnik2014}%
  \BibitemOpen
  \bibfield  {author} {\bibinfo {author} {\bibfnamefont {A.~R.}\ \bibnamefont
  {Mellnik}}, \bibinfo {author} {\bibfnamefont {J.~S.}\ \bibnamefont {Lee}},
  \bibinfo {author} {\bibfnamefont {A.}~\bibnamefont {Richardella}}, \bibinfo
  {author} {\bibfnamefont {J.~L.}\ \bibnamefont {Grab}}, \bibinfo {author}
  {\bibfnamefont {P.~J.}\ \bibnamefont {Mintun}}, \bibinfo {author}
  {\bibfnamefont {M.~H.}\ \bibnamefont {Fischer}}, \bibinfo {author}
  {\bibfnamefont {A.}~\bibnamefont {Vaezi}}, \bibinfo {author} {\bibfnamefont
  {A.}~\bibnamefont {Manchon}}, \bibinfo {author} {\bibfnamefont {E.-A.}\
  \bibnamefont {Kim}}, \bibinfo {author} {\bibfnamefont {N.}~\bibnamefont
  {Samarth}}, \ and\ \bibinfo {author} {\bibfnamefont {D.~C.}\ \bibnamefont
  {Ralph}},\ }\href {\doibase 10.1038/nature13534} {\bibfield  {journal}
  {\bibinfo  {journal} {Nature}\ }\textbf {\bibinfo {volume} {511}},\ \bibinfo
  {pages} {449} (\bibinfo {year} {2014})},\ \Eprint
  {http://arxiv.org/abs/1402.1124} {arXiv:1402.1124} \BibitemShut {NoStop}%
\bibitem [{\citenamefont {Wang}\ \emph {et~al.}(2015)\citenamefont {Wang},
  \citenamefont {Deorani}, \citenamefont {Banerjee}, \citenamefont {Koirala},
  \citenamefont {Brahlek}, \citenamefont {Oh},\ and\ \citenamefont
  {Yang}}]{Wang2015}%
  \BibitemOpen
  \bibfield  {author} {\bibinfo {author} {\bibfnamefont {Y.}~\bibnamefont
  {Wang}}, \bibinfo {author} {\bibfnamefont {P.}~\bibnamefont {Deorani}},
  \bibinfo {author} {\bibfnamefont {K.}~\bibnamefont {Banerjee}}, \bibinfo
  {author} {\bibfnamefont {N.}~\bibnamefont {Koirala}}, \bibinfo {author}
  {\bibfnamefont {M.}~\bibnamefont {Brahlek}}, \bibinfo {author} {\bibfnamefont
  {S.}~\bibnamefont {Oh}}, \ and\ \bibinfo {author} {\bibfnamefont
  {H.}~\bibnamefont {Yang}},\ }\href {\doibase 10.1103/PhysRevLett.114.257202}
  {\bibfield  {journal} {\bibinfo  {journal} {Phys. Rev. Lett.}\ }\textbf
  {\bibinfo {volume} {114}},\ \bibinfo {pages} {257202} (\bibinfo {year}
  {2015})},\ \Eprint {http://arxiv.org/abs/1505.07937} {arXiv:1505.07937}
  \BibitemShut {NoStop}%
\bibitem [{\citenamefont {Fan}\ \emph {et~al.}(2014)\citenamefont {Fan},
  \citenamefont {Upadhyaya}, \citenamefont {Kou}, \citenamefont {Lang},
  \citenamefont {Takei}, \citenamefont {Wang}, \citenamefont {Tang},
  \citenamefont {He}, \citenamefont {Chang}, \citenamefont {Montazeri},
  \citenamefont {Yu}, \citenamefont {Jiang}, \citenamefont {Nie}, \citenamefont
  {Schwartz}, \citenamefont {Tserkovnyak},\ and\ \citenamefont
  {Wang}}]{Fan2014}%
  \BibitemOpen
  \bibfield  {author} {\bibinfo {author} {\bibfnamefont {Y.}~\bibnamefont
  {Fan}}, \bibinfo {author} {\bibfnamefont {P.}~\bibnamefont {Upadhyaya}},
  \bibinfo {author} {\bibfnamefont {X.}~\bibnamefont {Kou}}, \bibinfo {author}
  {\bibfnamefont {M.}~\bibnamefont {Lang}}, \bibinfo {author} {\bibfnamefont
  {S.}~\bibnamefont {Takei}}, \bibinfo {author} {\bibfnamefont
  {Z.}~\bibnamefont {Wang}}, \bibinfo {author} {\bibfnamefont {J.}~\bibnamefont
  {Tang}}, \bibinfo {author} {\bibfnamefont {L.}~\bibnamefont {He}}, \bibinfo
  {author} {\bibfnamefont {L.-t.}\ \bibnamefont {Chang}}, \bibinfo {author}
  {\bibfnamefont {M.}~\bibnamefont {Montazeri}}, \bibinfo {author}
  {\bibfnamefont {G.}~\bibnamefont {Yu}}, \bibinfo {author} {\bibfnamefont
  {W.}~\bibnamefont {Jiang}}, \bibinfo {author} {\bibfnamefont
  {T.}~\bibnamefont {Nie}}, \bibinfo {author} {\bibfnamefont {R.~N.}\
  \bibnamefont {Schwartz}}, \bibinfo {author} {\bibfnamefont {Y.}~\bibnamefont
  {Tserkovnyak}}, \ and\ \bibinfo {author} {\bibfnamefont {K.~L.}\ \bibnamefont
  {Wang}},\ }\href {\doibase 10.1038/nmat3973} {\bibfield  {journal} {\bibinfo
  {journal} {Nat. Mater.}\ }\textbf {\bibinfo {volume} {13}},\ \bibinfo {pages}
  {699} (\bibinfo {year} {2014})}\BibitemShut {NoStop}%
\bibitem [{\citenamefont {Fan}\ \emph {et~al.}(2016{\natexlab{b}})\citenamefont
  {Fan}, \citenamefont {Kou}, \citenamefont {Upadhyaya}, \citenamefont {Shao},
  \citenamefont {Pan}, \citenamefont {Lang}, \citenamefont {Che}, \citenamefont
  {Tang}, \citenamefont {Montazeri}, \citenamefont {Murata}, \citenamefont
  {Chang}, \citenamefont {Akyol}, \citenamefont {Yu}, \citenamefont {Nie},
  \citenamefont {Wong}, \citenamefont {Liu}, \citenamefont {Wang},
  \citenamefont {Tserkovnyak},\ and\ \citenamefont {Wang}}]{Fan2015}%
  \BibitemOpen
  \bibfield  {author} {\bibinfo {author} {\bibfnamefont {Y.}~\bibnamefont
  {Fan}}, \bibinfo {author} {\bibfnamefont {X.}~\bibnamefont {Kou}}, \bibinfo
  {author} {\bibfnamefont {P.}~\bibnamefont {Upadhyaya}}, \bibinfo {author}
  {\bibfnamefont {Q.}~\bibnamefont {Shao}}, \bibinfo {author} {\bibfnamefont
  {L.}~\bibnamefont {Pan}}, \bibinfo {author} {\bibfnamefont {M.}~\bibnamefont
  {Lang}}, \bibinfo {author} {\bibfnamefont {X.}~\bibnamefont {Che}}, \bibinfo
  {author} {\bibfnamefont {J.}~\bibnamefont {Tang}}, \bibinfo {author}
  {\bibfnamefont {M.}~\bibnamefont {Montazeri}}, \bibinfo {author}
  {\bibfnamefont {K.}~\bibnamefont {Murata}}, \bibinfo {author} {\bibfnamefont
  {L.-T.}\ \bibnamefont {Chang}}, \bibinfo {author} {\bibfnamefont
  {M.}~\bibnamefont {Akyol}}, \bibinfo {author} {\bibfnamefont
  {G.}~\bibnamefont {Yu}}, \bibinfo {author} {\bibfnamefont {T.}~\bibnamefont
  {Nie}}, \bibinfo {author} {\bibfnamefont {K.~L.}\ \bibnamefont {Wong}},
  \bibinfo {author} {\bibfnamefont {J.}~\bibnamefont {Liu}}, \bibinfo {author}
  {\bibfnamefont {Y.}~\bibnamefont {Wang}}, \bibinfo {author} {\bibfnamefont
  {Y.}~\bibnamefont {Tserkovnyak}}, \ and\ \bibinfo {author} {\bibfnamefont
  {K.~L.}\ \bibnamefont {Wang}},\ }\href {\doibase 10.1038/nnano.2015.294}
  {\bibfield  {journal} {\bibinfo  {journal} {Nat. Nanotechnol.}\ }\textbf
  {\bibinfo {volume} {11}},\ \bibinfo {pages} {352} (\bibinfo {year}
  {2016}{\natexlab{b}})},\ \Eprint {http://arxiv.org/abs/1511.07442}
  {arXiv:1511.07442} \BibitemShut {NoStop}%
\bibitem [{\citenamefont {Yasuda}\ \emph {et~al.}(2017)\citenamefont {Yasuda},
  \citenamefont {Tsukazaki}, \citenamefont {Yoshimi}, \citenamefont {Kondou},
  \citenamefont {Takahashi}, \citenamefont {Otani}, \citenamefont {Kawasaki},\
  and\ \citenamefont {Tokura}}]{Yasuda2017}%
  \BibitemOpen
  \bibfield  {author} {\bibinfo {author} {\bibfnamefont {K.}~\bibnamefont
  {Yasuda}}, \bibinfo {author} {\bibfnamefont {A.}~\bibnamefont {Tsukazaki}},
  \bibinfo {author} {\bibfnamefont {R.}~\bibnamefont {Yoshimi}}, \bibinfo
  {author} {\bibfnamefont {K.}~\bibnamefont {Kondou}}, \bibinfo {author}
  {\bibfnamefont {K.~S.}\ \bibnamefont {Takahashi}}, \bibinfo {author}
  {\bibfnamefont {Y.}~\bibnamefont {Otani}}, \bibinfo {author} {\bibfnamefont
  {M.}~\bibnamefont {Kawasaki}}, \ and\ \bibinfo {author} {\bibfnamefont
  {Y.}~\bibnamefont {Tokura}},\ }\href {\doibase
  10.1103/PhysRevLett.119.137204} {\bibfield  {journal} {\bibinfo  {journal}
  {Phys. Rev. Lett.}\ }\textbf {\bibinfo {volume} {119}},\ \bibinfo {pages}
  {137204} (\bibinfo {year} {2017})}\BibitemShut {NoStop}%
\bibitem [{\citenamefont {Han}\ \emph {et~al.}(2017)\citenamefont {Han},
  \citenamefont {Richardella}, \citenamefont {Siddiqui}, \citenamefont
  {Finley}, \citenamefont {Samarth},\ and\ \citenamefont {Liu}}]{Han2017}%
  \BibitemOpen
  \bibfield  {author} {\bibinfo {author} {\bibfnamefont {J.}~\bibnamefont
  {Han}}, \bibinfo {author} {\bibfnamefont {A.}~\bibnamefont {Richardella}},
  \bibinfo {author} {\bibfnamefont {S.~A.}\ \bibnamefont {Siddiqui}}, \bibinfo
  {author} {\bibfnamefont {J.}~\bibnamefont {Finley}}, \bibinfo {author}
  {\bibfnamefont {N.}~\bibnamefont {Samarth}}, \ and\ \bibinfo {author}
  {\bibfnamefont {L.}~\bibnamefont {Liu}},\ }\href {\doibase
  10.1103/PhysRevLett.119.077702} {\bibfield  {journal} {\bibinfo  {journal}
  {Phys. Rev. Lett.}\ }\textbf {\bibinfo {volume} {119}},\ \bibinfo {pages}
  {077702} (\bibinfo {year} {2017})},\ \Eprint
  {http://arxiv.org/abs/1703.07470} {arXiv:1703.07470} \BibitemShut {NoStop}%
\bibitem [{\citenamefont {Wang}\ \emph {et~al.}(2017)\citenamefont {Wang},
  \citenamefont {Zhu}, \citenamefont {Wu}, \citenamefont {Yang}, \citenamefont
  {Yu}, \citenamefont {Ramaswamy}, \citenamefont {Mishra}, \citenamefont {Shi},
  \citenamefont {Elyasi}, \citenamefont {Teo}, \citenamefont {Wu},\ and\
  \citenamefont {Yang}}]{Wang2017d}%
  \BibitemOpen
  \bibfield  {author} {\bibinfo {author} {\bibfnamefont {Y.}~\bibnamefont
  {Wang}}, \bibinfo {author} {\bibfnamefont {D.}~\bibnamefont {Zhu}}, \bibinfo
  {author} {\bibfnamefont {Y.}~\bibnamefont {Wu}}, \bibinfo {author}
  {\bibfnamefont {Y.}~\bibnamefont {Yang}}, \bibinfo {author} {\bibfnamefont
  {J.}~\bibnamefont {Yu}}, \bibinfo {author} {\bibfnamefont {R.}~\bibnamefont
  {Ramaswamy}}, \bibinfo {author} {\bibfnamefont {R.}~\bibnamefont {Mishra}},
  \bibinfo {author} {\bibfnamefont {S.}~\bibnamefont {Shi}}, \bibinfo {author}
  {\bibfnamefont {M.}~\bibnamefont {Elyasi}}, \bibinfo {author} {\bibfnamefont
  {K.-l.}\ \bibnamefont {Teo}}, \bibinfo {author} {\bibfnamefont
  {Y.}~\bibnamefont {Wu}}, \ and\ \bibinfo {author} {\bibfnamefont
  {H.}~\bibnamefont {Yang}},\ }\href {\doibase 10.1038/s41467-017-01583-4}
  {\bibfield  {journal} {\bibinfo  {journal} {Nature Communications}\ }\textbf
  {\bibinfo {volume} {8}},\ \bibinfo {pages} {1364} (\bibinfo {year}
  {2017})}\BibitemShut {NoStop}%
\bibitem [{\citenamefont {Mahendra}\ \emph {et~al.}(2017)\citenamefont
  {Mahendra}, \citenamefont {Grassi}, \citenamefont {Chen}, \citenamefont
  {Jamali}, \citenamefont {{Reifsnyder Hickey}}, \citenamefont {Zhang},
  \citenamefont {Zhao}, \citenamefont {Li}, \citenamefont {Quarterman},
  \citenamefont {Lv}, \citenamefont {Li}, \citenamefont {Manchon},
  \citenamefont {Mkhoyan}, \citenamefont {Low},\ and\ \citenamefont
  {Wang}}]{Mahendra2017}%
  \BibitemOpen
  \bibfield  {author} {\bibinfo {author} {\bibfnamefont {D.}~\bibnamefont
  {Mahendra}}, \bibinfo {author} {\bibfnamefont {R.}~\bibnamefont {Grassi}},
  \bibinfo {author} {\bibfnamefont {J.-Y.}\ \bibnamefont {Chen}}, \bibinfo
  {author} {\bibfnamefont {M.}~\bibnamefont {Jamali}}, \bibinfo {author}
  {\bibfnamefont {D.}~\bibnamefont {{Reifsnyder Hickey}}}, \bibinfo {author}
  {\bibfnamefont {D.}~\bibnamefont {Zhang}}, \bibinfo {author} {\bibfnamefont
  {Z.}~\bibnamefont {Zhao}}, \bibinfo {author} {\bibfnamefont {H.}~\bibnamefont
  {Li}}, \bibinfo {author} {\bibfnamefont {P.}~\bibnamefont {Quarterman}},
  \bibinfo {author} {\bibfnamefont {Y.}~\bibnamefont {Lv}}, \bibinfo {author}
  {\bibfnamefont {M.}~\bibnamefont {Li}}, \bibinfo {author} {\bibfnamefont
  {A.}~\bibnamefont {Manchon}}, \bibinfo {author} {\bibfnamefont {K.~A.}\
  \bibnamefont {Mkhoyan}}, \bibinfo {author} {\bibfnamefont {T.}~\bibnamefont
  {Low}}, \ and\ \bibinfo {author} {\bibfnamefont {J.-P.}\ \bibnamefont
  {Wang}},\ }\href@noop {} {\bibfield  {journal} {\bibinfo  {journal}
  {arXiv:1703.03822}\ } (\bibinfo {year} {2017})}\BibitemShut {NoStop}%
\bibitem [{\citenamefont {Miron}\ \emph
  {et~al.}(2011{\natexlab{b}})\citenamefont {Miron}, \citenamefont {Garello},
  \citenamefont {Gaudin}, \citenamefont {Zermatten}, \citenamefont {Costache},
  \citenamefont {Auffret}, \citenamefont {Bandiera}, \citenamefont {Rodmacq},
  \citenamefont {Schuhl},\ and\ \citenamefont {Gambardella}}]{Miron2011}%
  \BibitemOpen
  \bibfield  {author} {\bibinfo {author} {\bibfnamefont {I.~M.}\ \bibnamefont
  {Miron}}, \bibinfo {author} {\bibfnamefont {K.}~\bibnamefont {Garello}},
  \bibinfo {author} {\bibfnamefont {G.}~\bibnamefont {Gaudin}}, \bibinfo
  {author} {\bibfnamefont {P.-J.}\ \bibnamefont {Zermatten}}, \bibinfo {author}
  {\bibfnamefont {M.~V.}\ \bibnamefont {Costache}}, \bibinfo {author}
  {\bibfnamefont {S.}~\bibnamefont {Auffret}}, \bibinfo {author} {\bibfnamefont
  {S.}~\bibnamefont {Bandiera}}, \bibinfo {author} {\bibfnamefont
  {B.}~\bibnamefont {Rodmacq}}, \bibinfo {author} {\bibfnamefont
  {A.}~\bibnamefont {Schuhl}}, \ and\ \bibinfo {author} {\bibfnamefont
  {P.}~\bibnamefont {Gambardella}},\ }\href {\doibase 10.1038/nature10309}
  {\bibfield  {journal} {\bibinfo  {journal} {Nature}\ }\textbf {\bibinfo
  {volume} {476}},\ \bibinfo {pages} {189} (\bibinfo {year}
  {2011}{\natexlab{b}})}\BibitemShut {NoStop}%
\bibitem [{\citenamefont {Garate}\ and\ \citenamefont
  {Franz}(2010)}]{Garate2010}%
  \BibitemOpen
  \bibfield  {author} {\bibinfo {author} {\bibfnamefont {I.}~\bibnamefont
  {Garate}}\ and\ \bibinfo {author} {\bibfnamefont {M.}~\bibnamefont {Franz}},\
  }\href {\doibase 10.1103/PhysRevLett.104.146802} {\bibfield  {journal}
  {\bibinfo  {journal} {Phys. Rev. Lett.}\ }\textbf {\bibinfo {volume} {104}},\
  \bibinfo {pages} {146802} (\bibinfo {year} {2010})},\ \Eprint
  {http://arxiv.org/abs/0911.0106} {arXiv:0911.0106} \BibitemShut {NoStop}%
\bibitem [{\citenamefont {Sakai}\ and\ \citenamefont
  {Kohno}(2014)}]{Sakai2014}%
  \BibitemOpen
  \bibfield  {author} {\bibinfo {author} {\bibfnamefont {A.}~\bibnamefont
  {Sakai}}\ and\ \bibinfo {author} {\bibfnamefont {H.}~\bibnamefont {Kohno}},\
  }\href {\doibase 10.1103/PhysRevB.89.165307} {\bibfield  {journal} {\bibinfo
  {journal} {Phys. Rev. B}\ }\textbf {\bibinfo {volume} {89}},\ \bibinfo
  {pages} {165307} (\bibinfo {year} {2014})},\ \Eprint
  {http://arxiv.org/abs/arXiv:1309.4195v1} {arXiv:arXiv:1309.4195v1}
  \BibitemShut {NoStop}%
\bibitem [{\citenamefont {Ho}\ \emph {et~al.}(2016)\citenamefont {Ho},
  \citenamefont {Wang}, \citenamefont {Siu}, \citenamefont {Yang},
  \citenamefont {Tan},\ and\ \citenamefont {Jalil}}]{Ho2016}%
  \BibitemOpen
  \bibfield  {author} {\bibinfo {author} {\bibfnamefont {C.~S.}\ \bibnamefont
  {Ho}}, \bibinfo {author} {\bibfnamefont {Y.}~\bibnamefont {Wang}}, \bibinfo
  {author} {\bibfnamefont {Z.~B.}\ \bibnamefont {Siu}}, \bibinfo {author}
  {\bibfnamefont {H.}~\bibnamefont {Yang}}, \bibinfo {author} {\bibfnamefont
  {S.~G.}\ \bibnamefont {Tan}}, \ and\ \bibinfo {author} {\bibfnamefont
  {M.~B.~A.}\ \bibnamefont {Jalil}},\ }\href {\doibase
  10.1038/s41598-017-00911-4} {\bibfield  {journal} {\bibinfo  {journal} {Sci.
  Rep.}\ ,\ \bibinfo {pages} {1}} (\bibinfo {year} {2016})},\ \Eprint
  {http://arxiv.org/abs/1611.08116} {arXiv:1611.08116} \BibitemShut {NoStop}%
\bibitem [{\citenamefont {Ndiaye}\ \emph {et~al.}(2017)\citenamefont {Ndiaye},
  \citenamefont {Akosa}, \citenamefont {Fischer}, \citenamefont {Vaezi},
  \citenamefont {Kim},\ and\ \citenamefont {Manchon}}]{Ndiaye2017}%
  \BibitemOpen
  \bibfield  {author} {\bibinfo {author} {\bibfnamefont {P.~B.}\ \bibnamefont
  {Ndiaye}}, \bibinfo {author} {\bibfnamefont {C.~A.}\ \bibnamefont {Akosa}},
  \bibinfo {author} {\bibfnamefont {M.~H.}\ \bibnamefont {Fischer}}, \bibinfo
  {author} {\bibfnamefont {A.}~\bibnamefont {Vaezi}}, \bibinfo {author}
  {\bibfnamefont {E.-A.}\ \bibnamefont {Kim}}, \ and\ \bibinfo {author}
  {\bibfnamefont {A.}~\bibnamefont {Manchon}},\ }\href {\doibase
  10.1103/PhysRevB.96.014408} {\bibfield  {journal} {\bibinfo  {journal} {Phys.
  Rev. B}\ }\textbf {\bibinfo {volume} {96}},\ \bibinfo {pages} {014408}
  (\bibinfo {year} {2017})},\ \Eprint {http://arxiv.org/abs/1509.06929}
  {arXiv:1509.06929} \BibitemShut {NoStop}%
\bibitem [{\citenamefont {Fischer}\ \emph {et~al.}(2016)\citenamefont
  {Fischer}, \citenamefont {Vaezi}, \citenamefont {Manchon},\ and\
  \citenamefont {Kim}}]{Fischer2016}%
  \BibitemOpen
  \bibfield  {author} {\bibinfo {author} {\bibfnamefont {M.~H.}\ \bibnamefont
  {Fischer}}, \bibinfo {author} {\bibfnamefont {A.}~\bibnamefont {Vaezi}},
  \bibinfo {author} {\bibfnamefont {A.}~\bibnamefont {Manchon}}, \ and\
  \bibinfo {author} {\bibfnamefont {E.-A.}\ \bibnamefont {Kim}},\ }\href
  {\doibase 10.1103/PhysRevB.93.125303} {\bibfield  {journal} {\bibinfo
  {journal} {Phys. Rev. B}\ }\textbf {\bibinfo {volume} {93}},\ \bibinfo
  {pages} {125303} (\bibinfo {year} {2016})},\ \Eprint
  {http://arxiv.org/abs/1305.1328} {arXiv:1305.1328} \BibitemShut {NoStop}%
\bibitem [{\citenamefont {Mahfouzi}\ \emph {et~al.}(2014)\citenamefont
  {Mahfouzi}, \citenamefont {Nagaosa},\ and\ \citenamefont
  {Nikoli{\'{c}}}}]{Mahfouzi2014}%
  \BibitemOpen
  \bibfield  {author} {\bibinfo {author} {\bibfnamefont {F.}~\bibnamefont
  {Mahfouzi}}, \bibinfo {author} {\bibfnamefont {N.}~\bibnamefont {Nagaosa}}, \
  and\ \bibinfo {author} {\bibfnamefont {B.~K.}\ \bibnamefont
  {Nikoli{\'{c}}}},\ }\href {\doibase 10.1103/PhysRevB.90.115432} {\bibfield
  {journal} {\bibinfo  {journal} {Phys. Rev. B}\ }\textbf {\bibinfo {volume}
  {90}},\ \bibinfo {pages} {115432} (\bibinfo {year} {2014})},\ \Eprint
  {http://arxiv.org/abs/1312.7091} {arXiv:1312.7091} \BibitemShut {NoStop}%
\bibitem [{\citenamefont {Mahfouzi}\ \emph {et~al.}(2016)\citenamefont
  {Mahfouzi}, \citenamefont {Nikoli{\'{c}}},\ and\ \citenamefont
  {Kioussis}}]{Mahfouzi2016}%
  \BibitemOpen
  \bibfield  {author} {\bibinfo {author} {\bibfnamefont {F.}~\bibnamefont
  {Mahfouzi}}, \bibinfo {author} {\bibfnamefont {B.~K.}\ \bibnamefont
  {Nikoli{\'{c}}}}, \ and\ \bibinfo {author} {\bibfnamefont {N.}~\bibnamefont
  {Kioussis}},\ }\href {\doibase 10.1103/PhysRevB.93.115419} {\bibfield
  {journal} {\bibinfo  {journal} {Phys. Rev. B}\ }\textbf {\bibinfo {volume}
  {93}},\ \bibinfo {pages} {115419} (\bibinfo {year} {2016})},\ \Eprint
  {http://arxiv.org/abs/1506.01303} {arXiv:1506.01303} \BibitemShut {NoStop}%
\bibitem [{\citenamefont {Fu}\ \emph {et~al.}(2007)\citenamefont {Fu},
  \citenamefont {Kane},\ and\ \citenamefont {Mele}}]{Fu2007}%
  \BibitemOpen
  \bibfield  {author} {\bibinfo {author} {\bibfnamefont {L.}~\bibnamefont
  {Fu}}, \bibinfo {author} {\bibfnamefont {C.~L.}\ \bibnamefont {Kane}}, \ and\
  \bibinfo {author} {\bibfnamefont {E.~J.}\ \bibnamefont {Mele}},\ }\href
  {\doibase 10.1103/PhysRevLett.98.106803} {\bibfield  {journal} {\bibinfo
  {journal} {Phys. Rev. Lett.}\ }\textbf {\bibinfo {volume} {98}},\ \bibinfo
  {pages} {106803} (\bibinfo {year} {2007})},\ \Eprint
  {http://arxiv.org/abs/0607699} {arXiv:0607699 [cond-mat]} \BibitemShut
  {NoStop}%
\bibitem [{\citenamefont {Roy}(2009)}]{Roy2009a}%
  \BibitemOpen
  \bibfield  {author} {\bibinfo {author} {\bibfnamefont {R.}~\bibnamefont
  {Roy}},\ }\href {\doibase 10.1103/PhysRevB.79.195322} {\bibfield  {journal}
  {\bibinfo  {journal} {Phys. Rev. B}\ }\textbf {\bibinfo {volume} {79}},\
  \bibinfo {pages} {195322} (\bibinfo {year} {2009})},\ \Eprint
  {http://arxiv.org/abs/0607531} {arXiv:0607531 [cond-mat]} \BibitemShut
  {NoStop}%
\bibitem [{\citenamefont {Peng}\ \emph {et~al.}(2016)\citenamefont {Peng},
  \citenamefont {Yang}, \citenamefont {Singh}, \citenamefont {Savrasov},\ and\
  \citenamefont {Yu}}]{Peng2016}%
  \BibitemOpen
  \bibfield  {author} {\bibinfo {author} {\bibfnamefont {X.}~\bibnamefont
  {Peng}}, \bibinfo {author} {\bibfnamefont {Y.}~\bibnamefont {Yang}}, \bibinfo
  {author} {\bibfnamefont {R.~R.}\ \bibnamefont {Singh}}, \bibinfo {author}
  {\bibfnamefont {S.~Y.}\ \bibnamefont {Savrasov}}, \ and\ \bibinfo {author}
  {\bibfnamefont {D.}~\bibnamefont {Yu}},\ }\href {\doibase
  10.1038/ncomms10878} {\bibfield  {journal} {\bibinfo  {journal} {Nat.
  Commun.}\ }\textbf {\bibinfo {volume} {7}},\ \bibinfo {pages} {10878}
  (\bibinfo {year} {2016})}\BibitemShut {NoStop}%
\bibitem [{\citenamefont {Marmolejo-Tejada}\ \emph {et~al.}(2017)\citenamefont
  {Marmolejo-Tejada}, \citenamefont {Dolui}, \citenamefont {Lazi{\'{c}}},
  \citenamefont {Chang}, \citenamefont {Smidstrup}, \citenamefont {Stradi},
  \citenamefont {Stokbro},\ and\ \citenamefont {Nikoli{\'{c}}}}]{Tejada2017}%
  \BibitemOpen
  \bibfield  {author} {\bibinfo {author} {\bibfnamefont {J.~M.}\ \bibnamefont
  {Marmolejo-Tejada}}, \bibinfo {author} {\bibfnamefont {K.}~\bibnamefont
  {Dolui}}, \bibinfo {author} {\bibfnamefont {P.}~\bibnamefont {Lazi{\'{c}}}},
  \bibinfo {author} {\bibfnamefont {P.-H.}\ \bibnamefont {Chang}}, \bibinfo
  {author} {\bibfnamefont {S.}~\bibnamefont {Smidstrup}}, \bibinfo {author}
  {\bibfnamefont {D.}~\bibnamefont {Stradi}}, \bibinfo {author} {\bibfnamefont
  {K.}~\bibnamefont {Stokbro}}, \ and\ \bibinfo {author} {\bibfnamefont
  {B.~K.}\ \bibnamefont {Nikoli{\'{c}}}},\ }\href {\doibase
  10.1021/acs.nanolett.7b02511} {\bibfield  {journal} {\bibinfo  {journal}
  {Nano Lett.}\ }\textbf {\bibinfo {volume} {17}},\ \bibinfo {pages} {5626}
  (\bibinfo {year} {2017})},\ \Eprint {http://arxiv.org/abs/1701.00462}
  {arXiv:1701.00462} \BibitemShut {NoStop}%
\bibitem [{\citenamefont {Zhang}\ \emph {et~al.}(2016)\citenamefont {Zhang},
  \citenamefont {Velev}, \citenamefont {Dang},\ and\ \citenamefont
  {Tsymbal}}]{Zhang2016}%
  \BibitemOpen
  \bibfield  {author} {\bibinfo {author} {\bibfnamefont {J.}~\bibnamefont
  {Zhang}}, \bibinfo {author} {\bibfnamefont {J.~P.}\ \bibnamefont {Velev}},
  \bibinfo {author} {\bibfnamefont {X.}~\bibnamefont {Dang}}, \ and\ \bibinfo
  {author} {\bibfnamefont {E.~Y.}\ \bibnamefont {Tsymbal}},\ }\href {\doibase
  10.1103/PhysRevB.94.014435} {\bibfield  {journal} {\bibinfo  {journal} {Phys.
  Rev. B}\ }\textbf {\bibinfo {volume} {94}},\ \bibinfo {pages} {014435}
  (\bibinfo {year} {2016})},\ \Eprint {http://arxiv.org/abs/1606.00763}
  {arXiv:1606.00763} \BibitemShut {NoStop}%
\bibitem [{\citenamefont {Hsu}\ \emph {et~al.}()\citenamefont {Hsu},
  \citenamefont {Park},\ and\ \citenamefont {Kim}}]{Hsu2017}%
  \BibitemOpen
  \bibfield  {author} {\bibinfo {author} {\bibfnamefont {Y.-T.}\ \bibnamefont
  {Hsu}}, \bibinfo {author} {\bibfnamefont {K.}~\bibnamefont {Park}}, \ and\
  \bibinfo {author} {\bibfnamefont {E.-A.}\ \bibnamefont {Kim}},\ }\href
  {http://arxiv.org/abs/1707.06319} {\ }\Eprint
  {http://arxiv.org/abs/1707.06319} {arXiv:1707.06319} \BibitemShut {NoStop}%
\bibitem [{\citenamefont {Marchand}\ and\ \citenamefont
  {Franz}(2012)}]{Marchand2012}%
  \BibitemOpen
  \bibfield  {author} {\bibinfo {author} {\bibfnamefont {D.~J.~J.}\
  \bibnamefont {Marchand}}\ and\ \bibinfo {author} {\bibfnamefont
  {M.}~\bibnamefont {Franz}},\ }\href {\doibase 10.1103/PhysRevB.86.155146}
  {\bibfield  {journal} {\bibinfo  {journal} {Phys. Rev. B}\ }\textbf {\bibinfo
  {volume} {86}},\ \bibinfo {pages} {155146} (\bibinfo {year}
  {2012})}\BibitemShut {NoStop}%
\bibitem [{\citenamefont {Hosur}\ \emph {et~al.}(2010)\citenamefont {Hosur},
  \citenamefont {Ryu},\ and\ \citenamefont {Vishwanath}}]{Hosur2010}%
  \BibitemOpen
  \bibfield  {author} {\bibinfo {author} {\bibfnamefont {P.}~\bibnamefont
  {Hosur}}, \bibinfo {author} {\bibfnamefont {S.}~\bibnamefont {Ryu}}, \ and\
  \bibinfo {author} {\bibfnamefont {A.}~\bibnamefont {Vishwanath}},\ }\href
  {\doibase 10.1103/PhysRevB.81.045120} {\bibfield  {journal} {\bibinfo
  {journal} {Phys. Rev. B}\ }\textbf {\bibinfo {volume} {81}},\ \bibinfo
  {pages} {045120} (\bibinfo {year} {2010})}\BibitemShut {NoStop}%
\bibitem [{\citenamefont {Wimmer}\ \emph {et~al.}(2016)\citenamefont {Wimmer},
  \citenamefont {Chadova}, \citenamefont {Seemann}, \citenamefont
  {K{\"{o}}dderitzsch},\ and\ \citenamefont {Ebert}}]{Wimmer2016}%
  \BibitemOpen
  \bibfield  {author} {\bibinfo {author} {\bibfnamefont {S.}~\bibnamefont
  {Wimmer}}, \bibinfo {author} {\bibfnamefont {K.}~\bibnamefont {Chadova}},
  \bibinfo {author} {\bibfnamefont {M.}~\bibnamefont {Seemann}}, \bibinfo
  {author} {\bibfnamefont {D.}~\bibnamefont {K{\"{o}}dderitzsch}}, \ and\
  \bibinfo {author} {\bibfnamefont {H.}~\bibnamefont {Ebert}},\ }\href
  {\doibase 10.1103/PhysRevB.94.054415} {\bibfield  {journal} {\bibinfo
  {journal} {Phys. Rev. B}\ }\textbf {\bibinfo {volume} {94}},\ \bibinfo
  {pages} {054415} (\bibinfo {year} {2016})},\ \Eprint
  {http://arxiv.org/abs/1604.02798} {arXiv:1604.02798} \BibitemShut {NoStop}%
\bibitem [{\citenamefont {Inoue}\ \emph {et~al.}(2004)\citenamefont {Inoue},
  \citenamefont {Bauer},\ and\ \citenamefont {Molenkamp}}]{Inoue2004}%
  \BibitemOpen
  \bibfield  {author} {\bibinfo {author} {\bibfnamefont {J.-i.}\ \bibnamefont
  {Inoue}}, \bibinfo {author} {\bibfnamefont {G.~E.~W.}\ \bibnamefont {Bauer}},
  \ and\ \bibinfo {author} {\bibfnamefont {L.~W.}\ \bibnamefont {Molenkamp}},\
  }\href {\doibase 10.1103/PhysRevB.70.041303} {\bibfield  {journal} {\bibinfo
  {journal} {Phys. Rev. B}\ }\textbf {\bibinfo {volume} {70}},\ \bibinfo
  {pages} {041303} (\bibinfo {year} {2004})},\ \Eprint
  {http://arxiv.org/abs/0402442} {arXiv:0402442 [cond-mat]} \BibitemShut
  {NoStop}%
\bibitem [{\citenamefont {Qaiumzadeh}\ \emph {et~al.}(2015)\citenamefont
  {Qaiumzadeh}, \citenamefont {Duine},\ and\ \citenamefont
  {Titov}}]{Qaiumzadeh2015}%
  \BibitemOpen
  \bibfield  {author} {\bibinfo {author} {\bibfnamefont {a.}~\bibnamefont
  {Qaiumzadeh}}, \bibinfo {author} {\bibfnamefont {R.~A.}\ \bibnamefont
  {Duine}}, \ and\ \bibinfo {author} {\bibfnamefont {M.}~\bibnamefont
  {Titov}},\ }\href {\doibase 10.1103/PhysRevB.92.014402} {\bibfield  {journal}
  {\bibinfo  {journal} {Phys. Rev. B}\ }\textbf {\bibinfo {volume} {92}},\
  \bibinfo {pages} {014402} (\bibinfo {year} {2015})},\ \Eprint
  {http://arxiv.org/abs/1503.06872} {arXiv:1503.06872} \BibitemShut {NoStop}%
\bibitem [{\citenamefont {Murakami}(2004)}]{Murakami2004b}%
  \BibitemOpen
  \bibfield  {author} {\bibinfo {author} {\bibfnamefont {S.}~\bibnamefont
  {Murakami}},\ }\href {\doibase 10.1103/PhysRevB.69.241202} {\bibfield
  {journal} {\bibinfo  {journal} {Phys. Rev. B}\ }\textbf {\bibinfo {volume}
  {69}},\ \bibinfo {pages} {241202} (\bibinfo {year} {2004})},\ \Eprint
  {http://arxiv.org/abs/0405001} {arXiv:0405001 [cond-mat]} \BibitemShut
  {NoStop}%
\bibitem [{\citenamefont {Krotkov}\ and\ \citenamefont {{Das
  Sarma}}(2006)}]{Krotkov2006}%
  \BibitemOpen
  \bibfield  {author} {\bibinfo {author} {\bibfnamefont {P.~L.}\ \bibnamefont
  {Krotkov}}\ and\ \bibinfo {author} {\bibfnamefont {S.}~\bibnamefont {{Das
  Sarma}}},\ }\href {\doibase 10.1103/PhysRevB.73.195307} {\bibfield  {journal}
  {\bibinfo  {journal} {Phys. Rev. B}\ }\textbf {\bibinfo {volume} {73}},\
  \bibinfo {pages} {195307} (\bibinfo {year} {2006})},\ \Eprint
  {http://arxiv.org/abs/0510114} {arXiv:0510114 [cond-mat]} \BibitemShut
  {NoStop}%
\bibitem [{\citenamefont {Sinova}\ \emph {et~al.}(2015)\citenamefont {Sinova},
  \citenamefont {Valenzuela}, \citenamefont {Wunderlich}, \citenamefont
  {Back},\ and\ \citenamefont {Jungwirth}}]{Sinova2015}%
  \BibitemOpen
  \bibfield  {author} {\bibinfo {author} {\bibfnamefont {J.}~\bibnamefont
  {Sinova}}, \bibinfo {author} {\bibfnamefont {S.~O.}\ \bibnamefont
  {Valenzuela}}, \bibinfo {author} {\bibfnamefont {J.}~\bibnamefont
  {Wunderlich}}, \bibinfo {author} {\bibfnamefont {C.~H.}\ \bibnamefont
  {Back}}, \ and\ \bibinfo {author} {\bibfnamefont {T.}~\bibnamefont
  {Jungwirth}},\ }\href {\doibase 10.1103/RevModPhys.87.1213} {\bibfield
  {journal} {\bibinfo  {journal} {Rev. Mod. Phys.}\ }\textbf {\bibinfo {volume}
  {87}},\ \bibinfo {pages} {1213} (\bibinfo {year} {2015})},\ \Eprint
  {http://arxiv.org/abs/1411.3249} {arXiv:1411.3249} \BibitemShut {NoStop}%
\bibitem [{\citenamefont {Dyrda{\l}}\ and\ \citenamefont
  {Barna{\'{s}}}(2015)}]{Dyrdal2015}%
  \BibitemOpen
  \bibfield  {author} {\bibinfo {author} {\bibfnamefont {A.}~\bibnamefont
  {Dyrda{\l}}}\ and\ \bibinfo {author} {\bibfnamefont {J.}~\bibnamefont
  {Barna{\'{s}}}},\ }\href {\doibase 10.1103/PhysRevB.92.165404} {\bibfield
  {journal} {\bibinfo  {journal} {Phys. Rev. B}\ }\textbf {\bibinfo {volume}
  {92}},\ \bibinfo {pages} {165404} (\bibinfo {year} {2015})},\ \Eprint
  {http://arxiv.org/abs/1505.02530v1} {arXiv:1505.02530v1} \BibitemShut
  {NoStop}%
\bibitem [{\citenamefont {Kim}\ \emph {et~al.}(2015)\citenamefont {Kim},
  \citenamefont {Lee}, \citenamefont {Lee},\ and\ \citenamefont
  {Stiles}}]{Kim2015}%
  \BibitemOpen
  \bibfield  {author} {\bibinfo {author} {\bibfnamefont {K.-W.}\ \bibnamefont
  {Kim}}, \bibinfo {author} {\bibfnamefont {K.-J.}\ \bibnamefont {Lee}},
  \bibinfo {author} {\bibfnamefont {H.-W.}\ \bibnamefont {Lee}}, \ and\
  \bibinfo {author} {\bibfnamefont {M.~D.}\ \bibnamefont {Stiles}},\ }\href
  {\doibase 10.1103/PhysRevB.92.224426} {\bibfield  {journal} {\bibinfo
  {journal} {Phys. Rev. B}\ }\textbf {\bibinfo {volume} {92}},\ \bibinfo
  {pages} {224426} (\bibinfo {year} {2015})},\ \Eprint
  {http://arxiv.org/abs/1412.6123} {arXiv:1412.6123} \BibitemShut {NoStop}%
\bibitem [{\citenamefont {Sinitsyn}\ \emph {et~al.}(2007)\citenamefont
  {Sinitsyn}, \citenamefont {MacDonald}, \citenamefont {Jungwirth},
  \citenamefont {Dugaev},\ and\ \citenamefont {Sinova}}]{Sinitsyn2007}%
  \BibitemOpen
  \bibfield  {author} {\bibinfo {author} {\bibfnamefont {N.~A.}\ \bibnamefont
  {Sinitsyn}}, \bibinfo {author} {\bibfnamefont {A.~H.}\ \bibnamefont
  {MacDonald}}, \bibinfo {author} {\bibfnamefont {T.}~\bibnamefont
  {Jungwirth}}, \bibinfo {author} {\bibfnamefont {V.~K.}\ \bibnamefont
  {Dugaev}}, \ and\ \bibinfo {author} {\bibfnamefont {J.}~\bibnamefont
  {Sinova}},\ }\href {\doibase 10.1103/PhysRevB.75.045315} {\bibfield
  {journal} {\bibinfo  {journal} {Phys. Rev. B}\ }\textbf {\bibinfo {volume}
  {75}},\ \bibinfo {pages} {045315} (\bibinfo {year} {2007})},\ \Eprint
  {http://arxiv.org/abs/0608682} {arXiv:0608682 [cond-mat]} \BibitemShut
  {NoStop}%
\bibitem [{\citenamefont {Isaev}\ \emph {et~al.}(2011)\citenamefont {Isaev},
  \citenamefont {Moon},\ and\ \citenamefont {Ortiz}}]{Isaev2011}%
  \BibitemOpen
  \bibfield  {author} {\bibinfo {author} {\bibfnamefont {L.}~\bibnamefont
  {Isaev}}, \bibinfo {author} {\bibfnamefont {Y.~H.}\ \bibnamefont {Moon}}, \
  and\ \bibinfo {author} {\bibfnamefont {G.}~\bibnamefont {Ortiz}},\ }\href
  {\doibase 10.1103/PhysRevB.84.075444} {\bibfield  {journal} {\bibinfo
  {journal} {Phys. Rev. B}\ }\textbf {\bibinfo {volume} {84}},\ \bibinfo
  {pages} {075444} (\bibinfo {year} {2011})},\ \Eprint
  {http://arxiv.org/abs/1103.0025} {arXiv:1103.0025} \BibitemShut {NoStop}%
\bibitem [{\citenamefont {Tanaka}\ \emph {et~al.}(2008)\citenamefont {Tanaka},
  \citenamefont {Kontani}, \citenamefont {Naito}, \citenamefont {Naito},
  \citenamefont {Hirashima}, \citenamefont {Yamada},\ and\ \citenamefont
  {Inoue}}]{Tanaka2008}%
  \BibitemOpen
  \bibfield  {author} {\bibinfo {author} {\bibfnamefont {T.}~\bibnamefont
  {Tanaka}}, \bibinfo {author} {\bibfnamefont {H.}~\bibnamefont {Kontani}},
  \bibinfo {author} {\bibfnamefont {M.}~\bibnamefont {Naito}}, \bibinfo
  {author} {\bibfnamefont {T.}~\bibnamefont {Naito}}, \bibinfo {author}
  {\bibfnamefont {D.~S.}\ \bibnamefont {Hirashima}}, \bibinfo {author}
  {\bibfnamefont {K.}~\bibnamefont {Yamada}}, \ and\ \bibinfo {author}
  {\bibfnamefont {J.}~\bibnamefont {Inoue}},\ }\href {\doibase
  10.1103/PhysRevB.77.165117} {\bibfield  {journal} {\bibinfo  {journal} {Phys.
  Rev. B}\ }\textbf {\bibinfo {volume} {77}},\ \bibinfo {pages} {165117}
  (\bibinfo {year} {2008})}\BibitemShut {NoStop}%
\bibitem [{\citenamefont {Sahin}\ and\ \citenamefont
  {Flatt{\'{e}}}(2015)}]{Sahin2015}%
  \BibitemOpen
  \bibfield  {author} {\bibinfo {author} {\bibfnamefont {C.}~\bibnamefont
  {Sahin}}\ and\ \bibinfo {author} {\bibfnamefont {M.~E.}\ \bibnamefont
  {Flatt{\'{e}}}},\ }\href {\doibase 10.1103/PhysRevLett.114.107201} {\bibfield
   {journal} {\bibinfo  {journal} {Phys. Rev. Lett.}\ }\textbf {\bibinfo
  {volume} {114}},\ \bibinfo {pages} {107201} (\bibinfo {year} {2015})},\
  \Eprint {http://arxiv.org/abs/1410.7319} {arXiv:1410.7319} \BibitemShut
  {NoStop}%
\bibitem [{\citenamefont {Chang}\ \emph {et~al.}(2015)\citenamefont {Chang},
  \citenamefont {Markussen}, \citenamefont {Smidstrup}, \citenamefont
  {Stokbro},\ and\ \citenamefont {Nikoli{\'{c}}}}]{Chang2015a}%
  \BibitemOpen
  \bibfield  {author} {\bibinfo {author} {\bibfnamefont {P.-H.}\ \bibnamefont
  {Chang}}, \bibinfo {author} {\bibfnamefont {T.}~\bibnamefont {Markussen}},
  \bibinfo {author} {\bibfnamefont {S.}~\bibnamefont {Smidstrup}}, \bibinfo
  {author} {\bibfnamefont {K.}~\bibnamefont {Stokbro}}, \ and\ \bibinfo
  {author} {\bibfnamefont {B.~K.}\ \bibnamefont {Nikoli{\'{c}}}},\ }\href
  {\doibase 10.1103/PhysRevB.92.201406} {\bibfield  {journal} {\bibinfo
  {journal} {Phys. Rev. B}\ }\textbf {\bibinfo {volume} {92}},\ \bibinfo
  {pages} {201406} (\bibinfo {year} {2015})},\ \Eprint
  {http://arxiv.org/abs/1503.08046} {arXiv:1503.08046} \BibitemShut {NoStop}%
\end{thebibliography}%
\end{document}